\documentclass{aa}
\usepackage[varg]{txfonts}
\usepackage{amsmath,amstext}

\usepackage{graphicx}
\usepackage{epstopdf}
\usepackage{float}
\usepackage{array}
\usepackage{mathtools}
\usepackage{booktabs}
\usepackage{subfigure}
\usepackage{url}
\usepackage{helvet}
\usepackage{tabularx}
\usepackage{multirow}
\usepackage{natbib}
\bibpunct{(}{)}{;}{a}{}{,} % to follow the A&A style
\begin{document}

\title{From dense hot Jupiter to low-density Neptune: The discovery of WASP-127b, WASP-136b, and WASP-138b}
\author{K. W. F. Lam\inst{\ref{Warwick}} \and F. Faedi\inst{\ref{Warwick}} \and D. J. A. Brown\inst{\ref{Warwick}} \and D. R. Anderson\inst{\ref{Keele}} \and L. Delrez\inst{\ref{Leige}} \and M. Gillon\inst{\ref{Leige}} \and G. H\'{e}brard\inst{\ref{IAP}, \ref{OHP}} \and M. Lendl\inst{\ref{SPI},\ref{Geneve}} \and L. Mancini\inst{\ref{MPIA}} \and J. Southworth\inst{\ref{Keele}}  \and B. Smalley \inst{\ref{Keele}} \and A. H. M. Triaud\inst{\ref{Geneve}, \ref{CPS}, \ref{Toronto}, \ref{IoA}} \and O. D. Turner \inst{\ref{Keele}} \and K. L. Hay\inst{\ref{StAndrew}} \and D. J. Armstrong\inst{\ref{Warwick},\ref{Belfast}} \and S. C. C. Barros\inst{\ref{Porto}, \ref{LAM}} \and A. S. Bonomo\inst{\ref{INAF}} \and F. Bouchy\inst{\ref{Geneve}, \ref{LAM}} \and P. Boumis\inst{\ref{Greece}} \and A. Collier Cameron\inst{\ref{StAndrew}} \and A. P. Doyle\inst{\ref{Warwick}} \and C. Hellier \inst{\ref{Keele}} \and T. Henning\inst{\ref{MPIA}} \and E. Jehin\inst{\ref{Leige}} \and G. King\inst{\ref{Warwick}} \and J. Kirk\inst{\ref{Warwick}} \and T. Louden\inst{\ref{Warwick}} \and P. F. L. Maxted\inst{\ref{Keele}} \and J. J. McCormac\inst{\ref{Warwick}} \and H. P. Osborn\inst{\ref{Warwick}} \and E. Palle\inst{\ref{IAC}} \and F. Pepe\inst{\ref{Geneve}} \and D.Pollacco\inst{\ref{Warwick}} \and J. Prieto-Arranz\inst{\ref{IAC}} \and D. Queloz\inst{\ref{Geneve},\ref{Cavendish}} \and J. Rey\inst{\ref{Geneve}} \and D. S\'{e}gransan\inst{\ref{Geneve}} \and S. Udry\inst{\ref{Geneve}} \and S. Walker\inst{\ref{Warwick}} \and R. G. West\inst{\ref{Warwick}} \and P. J. Wheatley\inst{\ref{Warwick}}}

\institute{University of Warwick, Department of Physics, Gibbet Hill Road, Coventry, CV4 7AL, UK \\
\email{w.f.lam@warwick.ac.uk}\label{Warwick} \and Astrophysics Group, Keele University, Staffordshire ST5 5BG, UK\label{Keele}  \and Institut d'Astrophysique et de Geophysique, Universite de Liege, Allee du 6 Aout, 17, Bat. B5C, Liege 1, Belgium\label{Leige} \and Institut d'Astrophysique de Paris, UMR7095 CNRS, Universit\'{e} Pierre \& Marie Curie, 98bis boulevard Arago, 75014 Paris, France\label{IAP}  \and Observatoire de Haute-Provence, Universit\'{e} d'Aix-Marseille \& CNRS, 04870 Saint Michel l'Observatoire, France\label{OHP} \and Space Research Institute, Austrian Academy of Sciences, Schmiedlstr. 6, 8042 Graz, Austria\label{SPI} \and Observatoire de Geneve, Universite de Geneve, 51 Chemin des Maillettes, 1290 Sauverny, Switzerland\label{Geneve}  \and Max Planck Institute for Astronomy, Koenigstuhl 17, D-69117, Heidelberg, Germany\label{MPIA} \and Centre for Planetary Sciences, University of Toronto at Scarborough, 1265 Military Trail, Toronto, ON, M1C 1A4, Canada\label{CPS}  \and 18 Department of Astronomy \& Astrophysics, University of Toronto, Toronto, ON M5S 3H4, Canada\label{Toronto}  \and 19 Institute of Astronomy, Madingley Road, Cambridge, CB3 0HA, United Kingdom\label{IoA} \\  \and SUPA, School of Physics and Astronomy, University of St. Andrews, North Haugh, Fife KY16 9SS, UK\label{StAndrew} \and ARC, School of Mathematics \& Physics, Queen’s University Belfast, University Road, Belfast BT7 1NN, UK\label{Belfast} \and Instituto de Astrof\'isica e Ci\^{e}ncias do Espa\c co, Universidade do Porto, CAUP, Rua das Estrelas, 4150-762 Porto, Portugal\label{Porto} \and Aix Marseille Universit\'e, CNRS, LAM (Laboratoire d'Astrophysique de Marseille) UMR 7326, 13388 Marseille, France\label{LAM} \and INAF - Osservatorio Astrofisico di Torino, Via Osservatorio 20, I-10025, Pino Torinese, Italy\label{INAF} \and Institute for Astronomy, Astrophysics, Space Applications and Remote Sensing, National Observatory of Athens, 15236 Penteli, Greece\label{Greece} \and Instituto de Astrofisica de Canarias, Via Lactea sn, 38200, La Laguna, Tenerife, Spain\label{IAC} \and Cavendish Laboratory, J J Thomson Avenue, Cambridge, CB3 0HE, UK\label{Cavendish} \\
}

\date{Received date / Accepted date }

\abstract {We report three newly discovered exoplanets from the SuperWASP survey. WASP-127b is a heavily inflated super-Neptune of mass $0.18 \pm 0.02~\rm M_J$ and radius $1.37 \pm 0.04~\rm R_J$. This is one of the least massive planets discovered by the WASP project. It orbits a bright host star ($\rm V_{mag} = 10.16$) of spectral type G5 with a period of $4.17$ days. WASP-127b is a low-density planet that has an extended atmosphere with a scale height of $2500 \pm 400~\rm km$, making it an ideal candidate for transmission spectroscopy. WASP-136b and WASP-138b are both hot Jupiters with mass and radii of $1.51 \pm 0.08~\rm M_J$ and $1.38 \pm 0.16~\rm R_J$, and $1.22 \pm 0.08~\rm M_J$ and $1.09 \pm 0.05~\rm R_J$, respectively. WASP-136b is in a $5.22$-day orbit around an F9 subgiant star with a mass of $1.41 \pm 0.07~\rm M_{\odot}$ and a radius of $2.21 \pm 0.22~\rm R_{\odot}$. The discovery of WASP-136b could help constrain the characteristics of the giant planet population around evolved stars. WASP-138b orbits an F7 star with a period of $3.63$ days. Its radius agrees with theoretical values from standard models, suggesting the presence of a heavy element core with a mass of $\sim 10 ~\rm M_{\oplus}$. The discovery of these new planets helps in exploring the diverse compositional range of short-period planets, and will aid our understanding of the physical characteristics of both gas giants and low-density planets.}


\titlerunning{Discovery of WASP-127b, WASP-136b and WASP-138b}
\authorrunning{Lam et al.}

\maketitle

\section{Introduction}
\hspace{0.5cm}Over 3000 exoplanets have been discovered as of 2016 July.\footnote{\url{http://exoplanetarchive.ipac.caltech.edu/}} The space-based mission \textit{Kepler} \citep{2010Sci...327..977B} and its successor \textit{K2} \citep{2014PASP..126..398H} have discovered a large number of transiting planets. The results show that small Earth- and Neptune-sized planets are common around solar-like stars \citep{2011ApJ...736...19B}. The {\it Kepler} discoveries have provided a large sample of planetary systems that are very useful for statistical studies of planetary populations (e.g. \citealt{2013ApJ...766...81F,2013ApJ...767...95D,2013PNAS..11019273P}). However, most of the {\it Kepler} planets orbit very faint stars, for which it is quite difficult to obtain precise radial velocity (RV) measurements that are necessary to constrain planetary masses. On the other hand, ground-based systematic surveys (e.g. HATNet: \citealt{2002PASP..114..974B}; SuperWASP: \citealt{2006PASP..118.1407P}; KELT: \citealt{2007PASP..119..923P}; QES: \citealt{2013AcA....63..465A}; HATSouth: \citet{2013PASP..125..154B}; NGTS: \citealt{2013EPSC....8..234W}) usually provide a large number of candidates around stars that are sufficiently bright to enable measuring their RVs, from which strong observational constraints for theoretical studies can be obtained. The upcoming space mission, TESS \citep{2015JATIS...1a4003R}, will target main-sequence dwarf stars that are brighter than $13^{\rm th}$ magnitude and could yield over 1000 planets smaller than Neptune. In addition, all-sky surveys such as MASCARA \citep{2012SPIE.8444E..0IS}, Evryscope, \citep{2015PASP..127..234L} and Fly's Eye Camera \citep{2016PASP..128d5002P} will also be able to provide planet candidates around bright stars for detailed characterisation.

Precise measurements and analysis of these systems reveals that planets with similar masses can have very different radii, resulting in very unique physical characteristics. For example, Jupiter-mass planets can range in radii from $0.775 \rm R_J$ (WASP-59b \citealt{2013A&A...549A.134H}) to $1.932 \rm R_J$ (WASP-17b \citealt{2010ApJ...709..159A,2012MNRAS.426.1338S}). A growing number of short-period sub-Saturn and super-Neptune mass planets such as WASP-39b ($\rm M_p = 0.28 M_J$; \citealt{2011A&A...531A..40F}), HAT-P-11b ($\rm M_p = 0.081 M_J$; \citealt{2010ApJ...710.1724B}), and HAT-P-26b ($\rm M_p = 0.059 M_J$; \citealt{2011ApJ...728..138H}) were also found. Many of these planets were found to have radii larger than predicted from standard coreless models (e.g.~\citealt{2007ApJ...659.1661F}). Some theories suggest that the planet radius is correlated with the equilibrium temperature. For example, strong stellar irradiation could heat up the planet, inflating its radius \citep{1996ApJ...459L..35G}. The planetary interior could also be tidally heated as the orbit circularises \citep{2001ApJ...548..466B,2003ApJ...592..555B}. An enhanced atmospheric opacity can hinder the cooling process of the planet such that the planet radius can remain larger for longer \citep{2007ApJ...661..502B}. The interaction between stellar wind and the magnetic field of the planet can lead to Ohmic heating, which could also influence the temperature of the planet \citep{2011ApJ...738....1B}. Low-density planets with extended atmospheres orbiting bright host stars are ideal targets for transmission spectroscopy, which can further reveal the composition of planets. 
%some of the highest-detection strength transmission spectra we have are for extended objects such as WASP-103, WASP-47, WASP-17, etc. and that is purely exist because of their extended atmospheres.

We present here the discovery of three new planets, WASP-127b, WASP-136b, and WASP-138b, discovered by the SuperWASP survey. Section \ref{observation} summarises the observations from the WASP detection, follow-up photometry, and spectroscopic data of each of the planets. In Sect.~\ref{results}, we describe our analysis and present the derived results of the system parameters. Lastly, we discuss the system characteristics and how evolution theories could explain the existence of these planets in Sect.~\ref{discussion}.

\section{\label{observation}Observations}

		\subsection{SuperWASP}
		\hspace{0.5cm}The SuperWASP-North facility is situated at the Observatorio del Roque de los Muchachos in La Palma, Canary Islands, and the SuperWASP-South facility is located at the Sutherland Station of the South African Astronomical Observatory. Both cameras consist of an array of 8 Canon 200mm, f/1.8 telephoto lenses that are linked to e2v CCDs of $2048 \times 2048$ pixels each. The cameras in each of the facilities provide a total field of view of $8\times64$ square degrees and a pixel scale of 13.7" \citep{2006PASP..118.1407P}. 
		
		WASP data were observed with an exposure time of 30 seconds and an average cadence of 8-10 minutes. The data were reduced with the pipeline as described in \citet{2006PASP..118.1407P}. The box least-squares fit \citep{2002A&A...391..369K} and the SysRem detrending algorithm \citep{2005MNRAS.356.1466T} were used to analyse light curves from multiple seasons to determine planetary transit signals and provide system parameters for the candidates \citep{2006MNRAS.373..799C}. 
		
		WASP-127 was observed between 2006 July 03 and 2014 June 13. A total number of 87349 photometric data points were taken. The number of transits and in-transit data depends on the camera used and the season in which the star was observed. In the best season, 57 transits were observed with 1084 in-transit data points. There were always more than 500 in-transit data points obtained in all seasons, with more than 6 transits observed.
		WASP-136 was observed between 2006 December 29 and 2014 December 06. A total number of 97827 photometric data points were taken. In the best season, 44 transits were observed with 2253 in-transit data points. Except for a few seasons, there were always more than 500 in-transit data points, with more than 7 transits observed.
		WASP-138 was observed between 2008 April 02 and 2012 December 20. A total number of 39381 photometric data points were taken. In the best season, 58 transits were observed with 1077 in-transit data points. There were always more than 87 in-transit data points, with more than 7 transits observed.
		
		Each set of data was analysed with the transit search algorithm \citep{2007MNRAS.380.1230C} and was flagged as planetary candidates for follow-up observations. The algorithm revealed that WASP-127 has a periodicity of $P = 4.18$ days, transit duration of $T_{14} \approx 3.6$ hours, and a depth of $\sim5.8$ mmag. WASP-136 showed a period of $P = 5.22$ days, transit duration of $T_{14} \approx 5.2$ hours, and a depth of $\sim2.9$ mmag. The transit signal of WASP-138 showed a period of $P = 3.6$ days, transit duration of $T_{14} \approx 4.1$ hours, and a depth of $\sim8.2$ mmag. Subsequent follow-up photometry and spectroscopy were obtained to confirm the existence of the planets and to characterise their physical parameters.

		\begin{table*}[!t]
		\caption{\label{Photometric}Photometric properties of WASP-127, WASP-136, and WASP-138.}
		\footnotesize
		\centering
		\begin{tabular}{cccc}
		\hline 
		Parameter & WASP-127 & WASP-136 & WASP-138 \\ 
	\hline 
	\hline
		Identifier & 1SWASP J104214.08$-$035006.3 & 1SWASP J000118.17$-$085534.6 & 1SWASP J024633.37$-$002750.0 \\
%		\hline 
		RA(J2000) & 10:42:14.08 & 00:01:18.17 & 02:46:33.37 \\ 
%		\hline
		Dec(J2000) & $-$03:50:06.3 & $-$08:55:34.6 & $-$00:27:50.0 \\ 
%		\hline 
		$B$ & $10.79$ & $10.39$ & $12.28$ \\ 
%		\hline 
		$V$ & $10.15$  &$9.98$  & $11.81$ \\ 
%		\hline 
		$R$ & $9.74$ & $9.71$ & $11.40$ \\ 
%		\hline 
		$H$ & $8.74$ & $8.79$ & $10.54$ \\ 
%		\hline 
		$K$ & $8.64$ & $8.81$ & $10.49$ \\ 
		\hline 
		\end{tabular}
		\end{table*}

		\subsection{Spectroscopic follow-up}
		\hspace{0.5cm}We have obtained spectra of WASP-127, WASP-136, and WASP-138 with the SOPHIE and CORALIE spectrographs. SOPHIE is mounted on the $1.93$ m telescope \citep{2009A&A...505..853B} on the Observatoire de Haute-Provence (OHP) and CORALIE is mounted on the $1.2$ m Euler-Swiss telescope \citep{2000A&A...359L..13Q,2002A&A...388..632P} in La Silla, Chile. We obtained the SOPHIE observations in high-efficiency mode ($R=40000$), while CORALIE observations were made with an instrumental resolution of $R=55000$.

\begin{figure}[htbp!]
	    \includegraphics[width=0.5\textwidth]{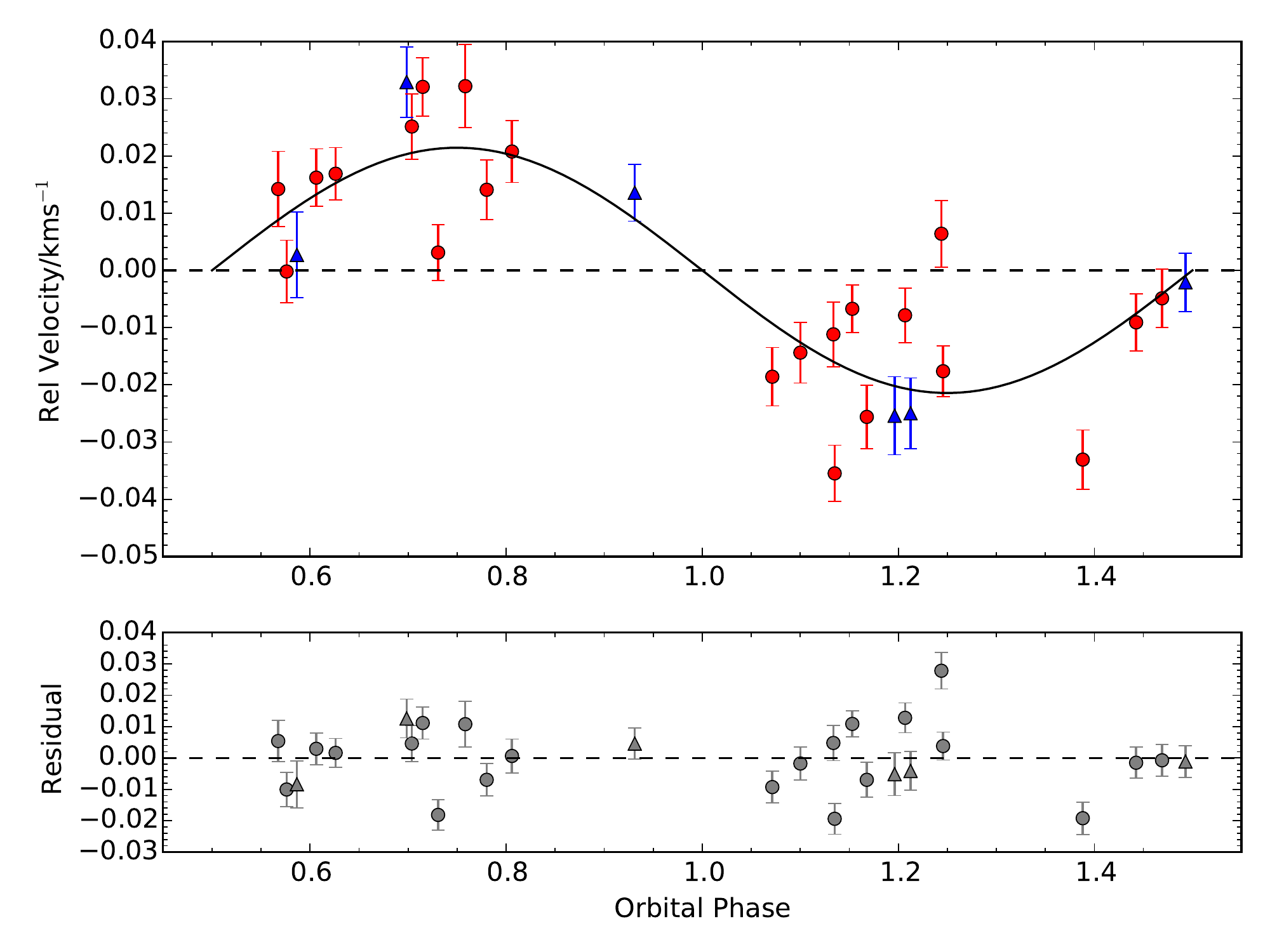}
	    \caption{\label{RV127_wosophie} \textit{Upper panel}: Phase-folded radial velocity of WASP-127 as a function of the orbital phase with best-fit RV curve (from the analysis with only the CORALIE RVs) plotted as a black solid line. CORALIE data observed before the instrumental upgrade are denoted by red circles, while data taken after the upgrade are denoted by blue triangles. \textit{Lower panel}: Residuals from the RV fit as a function of orbital phase.}
	\end{figure}

\begin{figure}[htbp!]
	    \includegraphics[width=0.5\textwidth]{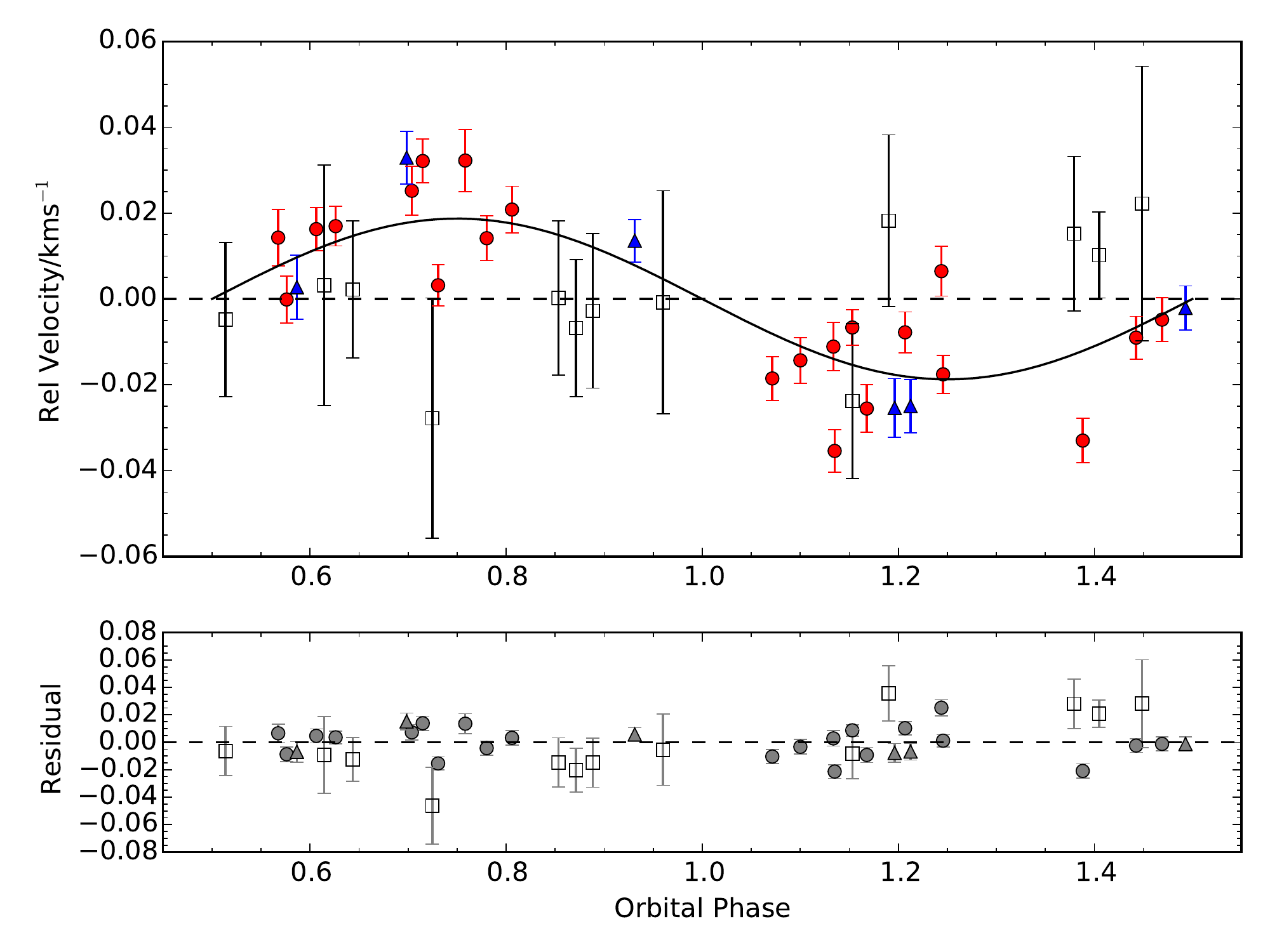}
	    \caption{\label{RV127_wsophie} Same as Figure \ref{RV127_wosophie} with SOPHIE data included as black open squares. The best-fit RV curve is obtained from our solution with both CORALIE and SOPHIE RVs.}
	\end{figure}

\begin{figure}[htbp!]
	    \includegraphics[width=0.5\textwidth]{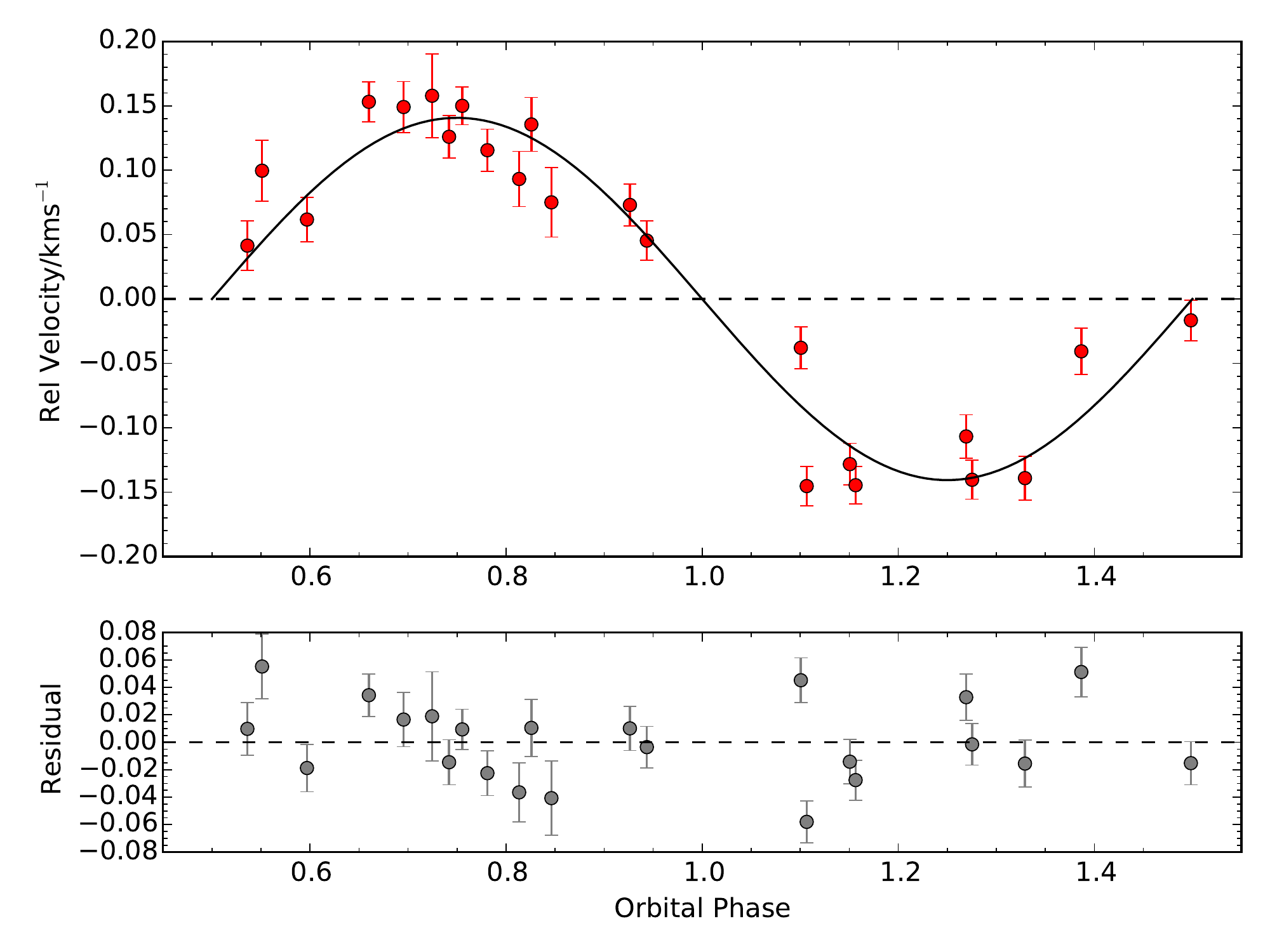}
	    \caption{\label{RV136}\textit{Upper panel}: Phase-folded radial velocity of WASP-136, measured by CORALIE (red circles), as a function of the orbital phase with the best-fit RV curve plotted as a black solid line. \textit{Lower panel}: Residuals from the RV fit as a function of orbital phase.}
	\end{figure}

\begin{figure}[htbp!]
	    \includegraphics[width=0.5\textwidth]{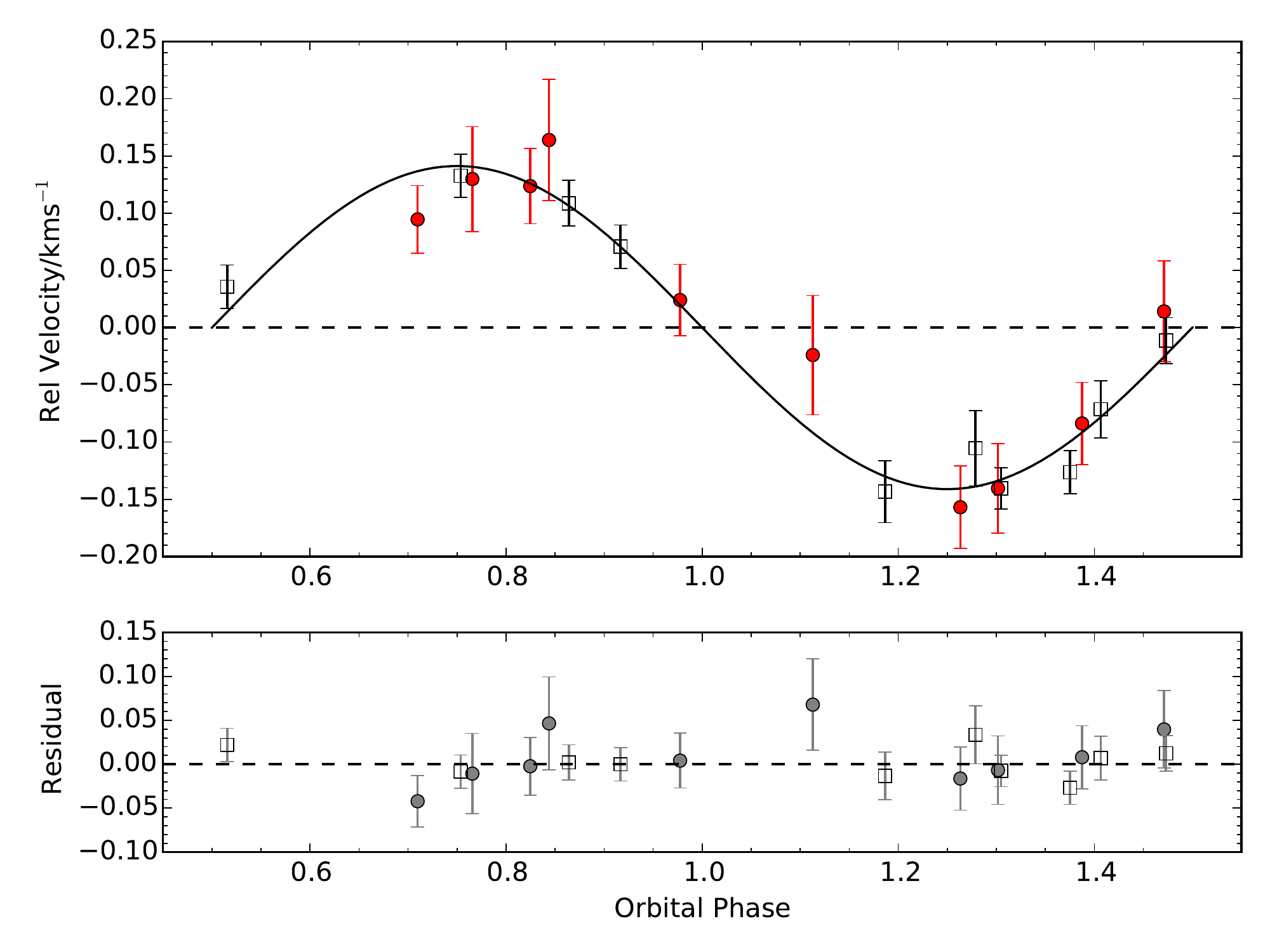}
	    \caption{\label{RV138}\textit{Upper panel}: Phase-folded radial velocity of WASP-138 as a function of the orbital phase with the best-fit RV curve plotted as a black solid line. CORALIE data are denoted by red circles and SOPHIE data are represented by black open squares. \textit{Lower panel}: Residuals from the RV fit as a function of orbital phase.}
	\end{figure}

A total of 28 spectral measurements of WASP-127 were taken between 2013 April 18 and 2015 April 9 using CORALIE. Six of the measurements were obtained after the instrumental upgrade in November 2014, which led to an offset of the zero point of the instrument, hence they will be treated as if they were from different instruments in our analysis. In addition, 13 SOPHIE measurements of WASP-127 were obtained between 2013 April 18 and 2014 December 31. Twenty-three CORALIE spectra of WASP-136 were obtained from 2014 June 24 to 2014 October 28. Between 2014 October 20 and 2015 January 25, 10 SOPHIE and 10 CORALIE measurements were made for WASP-138. All SOPHIE and CORALIE data were reduced with their respective standard reduction pipelines. The data were calibrated against radial velocity (RV) standards, yielding absolute RV measurements \citep{1996A&AS..119..373B}. We computed the RV of each system with a weighted cross-correlation method as described in \citet{1996A&AS..119..373B} and \citet{2002A&A...388..632P}. Figures \ref{RV127_wosophie} and \ref{RV127_wsophie} show the phase-folded RV measurements of WASP-127 from two different analyses (see Sect.~\ref{mcmcanalysis}). Similar RV plots of WASP-136 and WASP-138 are shown in Figs.~\ref{RV136} and \ref{RV138}, respectively. All CORALIE data observed before the instrumental upgrade are represented by red circles, and blue triangles are data obtained after the upgrade. SOPHIE measurements are denoted by open black squares.

\begin{figure}[htbp!]
	    \includegraphics[width=0.55\textwidth]{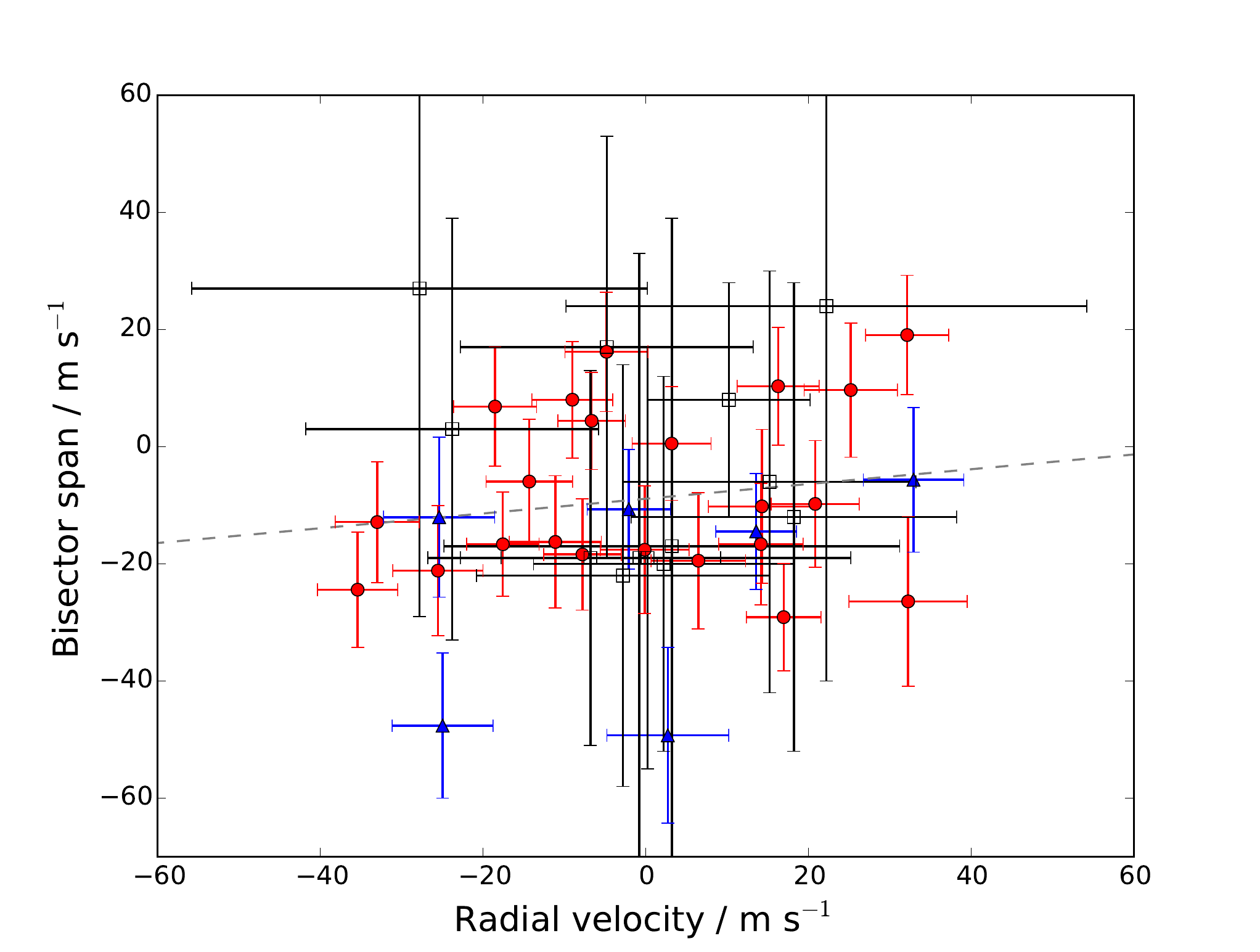}
	    \caption{\label{BIS127}Radial velocity bisector span of WASP-127 as a function of the relative radial velocity. CORALIE data before and after the instrumental upgrade are represented by red circles and blue triangles, respectively. SOPHIE data are denoted by open black squares. The grey dashed line shows the linear best fit to the data.}
	\end{figure}

WASP-127 has a visual companion located 41" away. We inspected the line bisector and searched for asymmetry in the stellar line profiles that may have arisen through stellar activity or a blended binary system \citep{2001A&A...379..279Q}. Figure \ref{BIS127} shows the bisector velocity span (V$\rm _{span}$) as a function of RV. No correlation is seen between V$\rm _{span}$ and RV, supporting the detection of a genuine planetary signal. Similar results were found for WASP-136 and WASP-138, where no correlation is seen between $\rm V_{span}$ and RV.

	\begin{figure}[htbp!]
	    \includegraphics[width=0.55\textwidth]{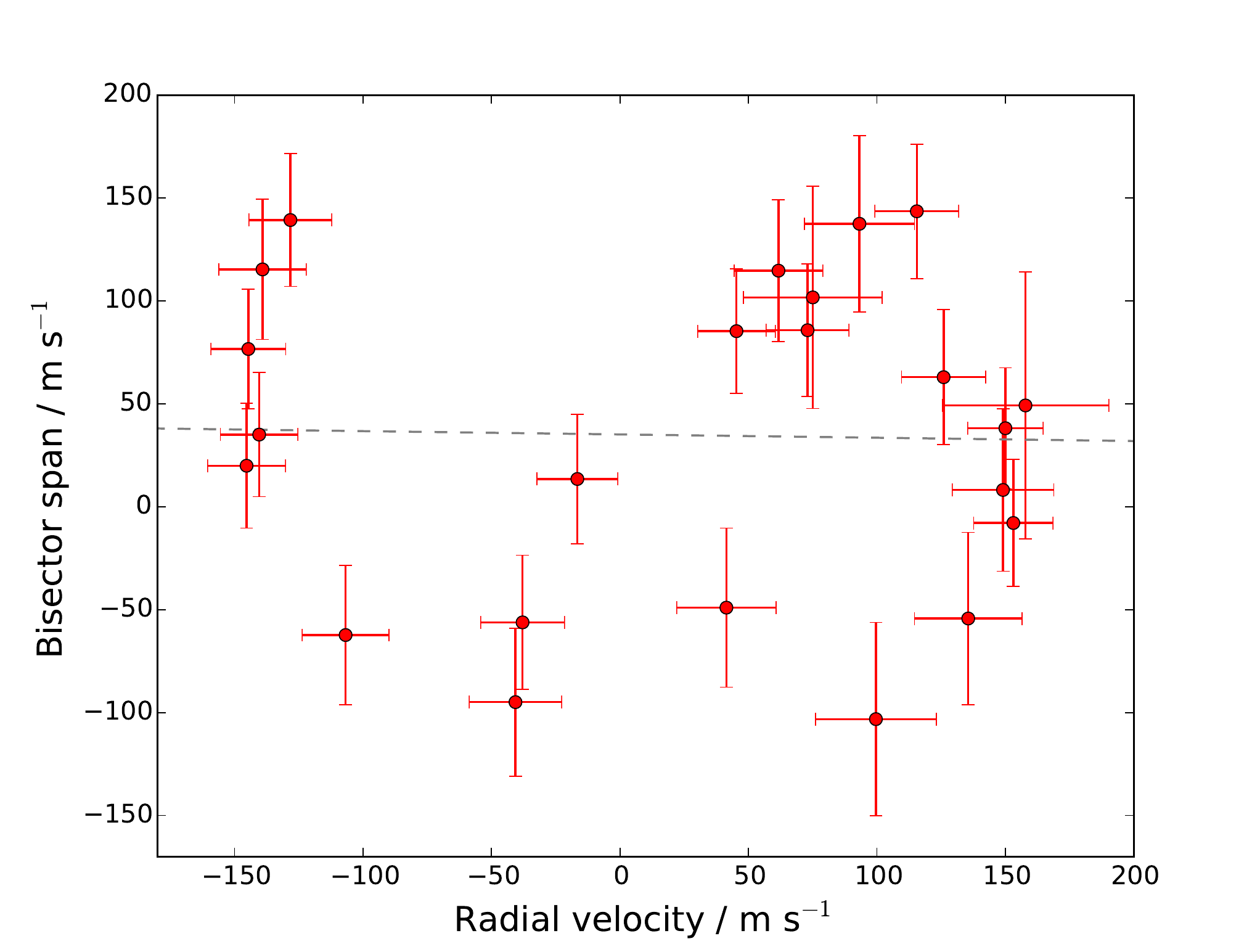}
	    \caption{\label{BIS136}Radial velocity bisector span of WASP-136 as a function of the relative radial velocity. The grey dashed line shows the linear best fit to the data.}
	\end{figure}	

	\begin{figure}[htbp!]
	    \includegraphics[width=0.55\textwidth]{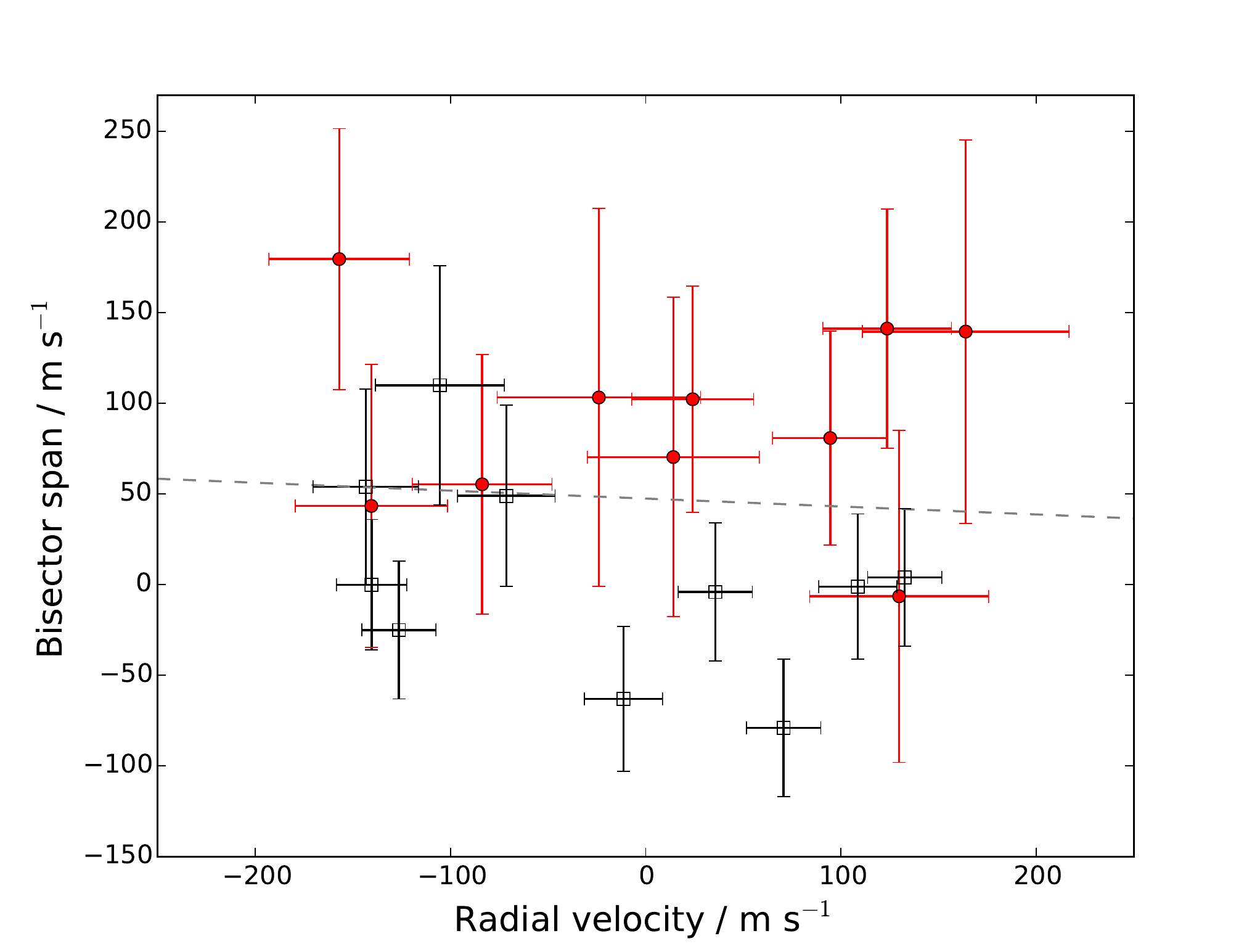}
	    \caption{\label{BIS138}Radial velocity bisector span of WASP-138 as a function of the relative radial velocity. CORALIE data are denoted by red circles and SOPHIE data are represented by open black squares. The grey dashed line shows the linear best fit to the data.}
	\end{figure}

		\subsection{Photometric follow-up}
	\begin{table}[h!]
	\centering
			\tiny
		\caption{\label{PhotometrySummary}Photometry observation of WASP-127, WASP-136, and WASP-138.}
\begin{tabular}{lllll}
\hline
Planet & Date & Instrument & Filter & Comment \\ 
\hline 
\multirow{4}{*}{WASP-127b} & \multicolumn{1}{l}{18/03/2014} & \multicolumn{1}{l}{TRAPPIST} & \multicolumn{1}{l}{z} & \multicolumn{1}{l}{partial transit} \\
										 & \multicolumn{1}{l}{28/04/2014} & \multicolumn{1}{l}{EulerCam} & \multicolumn{1}{l}{Gunn r} & \multicolumn{1}{l}{partial transit} \\
										 & \multicolumn{1}{l}{13/02/2016} & \multicolumn{1}{l}{LT RISE} & \multicolumn{1}{l}{V + R} & \multicolumn{1}{l}{full transit} \\ 
										 & \multicolumn{1}{l}{18/04/2016} & \multicolumn{1}{l}{Zeiss 1.23m} & \multicolumn{1}{l}{Cousins-I} & \multicolumn{1}{l}{full transit} \\ 
\hline 
\multirow{3}{*}{WASP-136b} & \multicolumn{1}{l}{21/08/2014} & \multicolumn{1}{l}{EulerCam} & \multicolumn{1}{l}{z} & \multicolumn{1}{l}{partial transit} \\
										 & \multicolumn{1}{l}{24/11/2014} & \multicolumn{1}{l}{TRAPPIST} & \multicolumn{1}{l}{z} & \multicolumn{1}{l}{partial transit} \\
										 & \multicolumn{1}{l}{21/08/2015} & \multicolumn{1}{l}{EulerCam} & \multicolumn{1}{l}{z} & \multicolumn{1}{l}{full transit} \\
\hline 
\multirow{1}{*}{WASP-138b} & \multicolumn{1}{l}{17/12/2015} & \multicolumn{1}{l}{EulerCam} & \multicolumn{1}{l}{NGTS} & \multicolumn{1}{l}{partial transit} \\
%\hline 
\hline 
\end{tabular} 	
\end{table}			
		
\begin{figure}[htbp!]
	\centering
	    \includegraphics[scale=0.45]{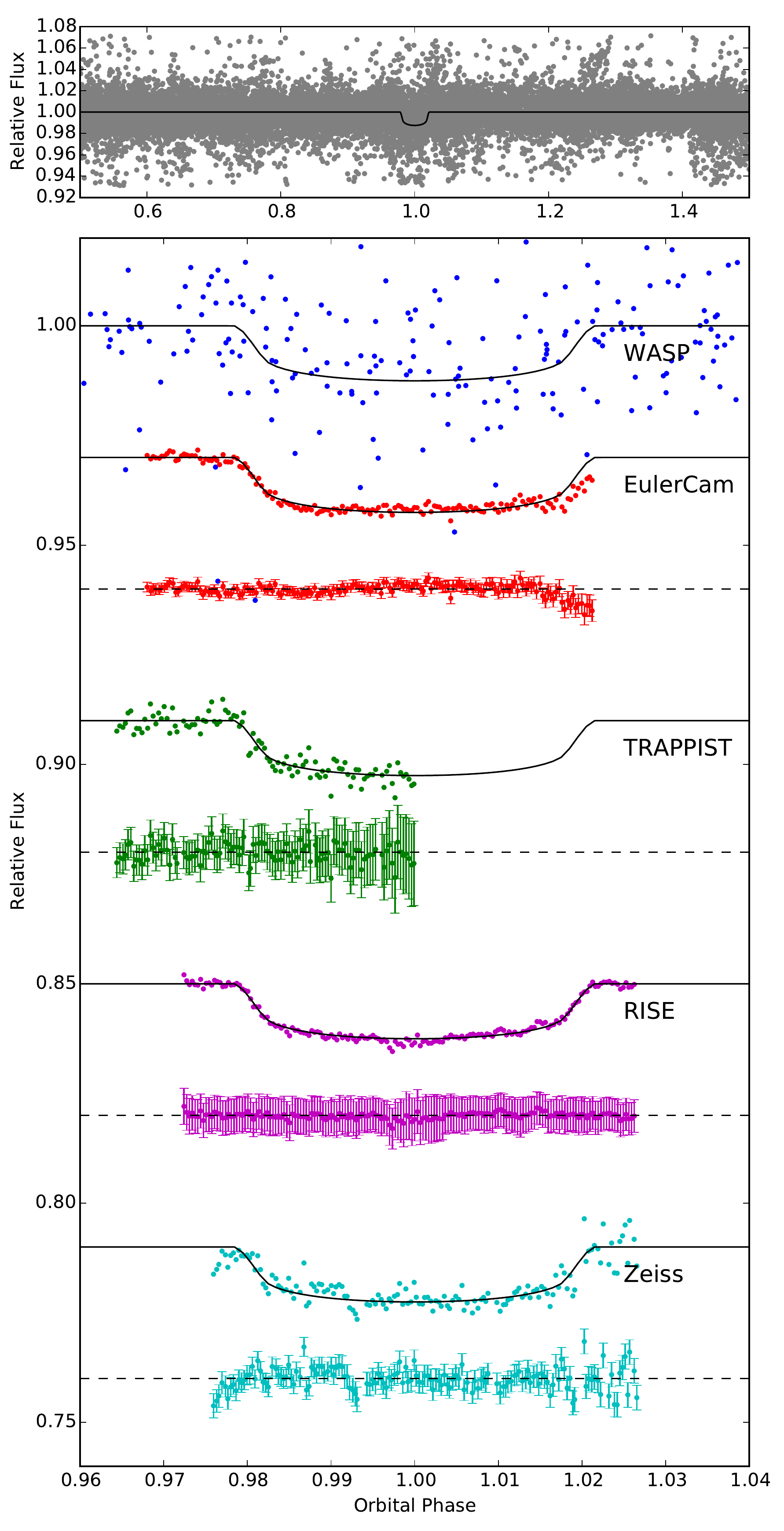}
	    \caption{\label{photomtery127}Photometry follow-up of WASP-127 observed from EULERCam (red), TRAPPIST(green), RISE(magenta) and Zeiss(cyan). The data are phase-folded with the ephemeris from our analysis. The light curves are assigned an arbitrary offset from the zero magnitute and are binned to a 2-minute cadence for clarity. The best-fit transit model from \citet{2002ApJ...580L.171M} is plotted as a black solid line, and the residuals of the fit are plotted directly below the light curves.}
	\end{figure}

\begin{figure}[htbp!]
	\centering
		    \includegraphics[width=0.45\textwidth]{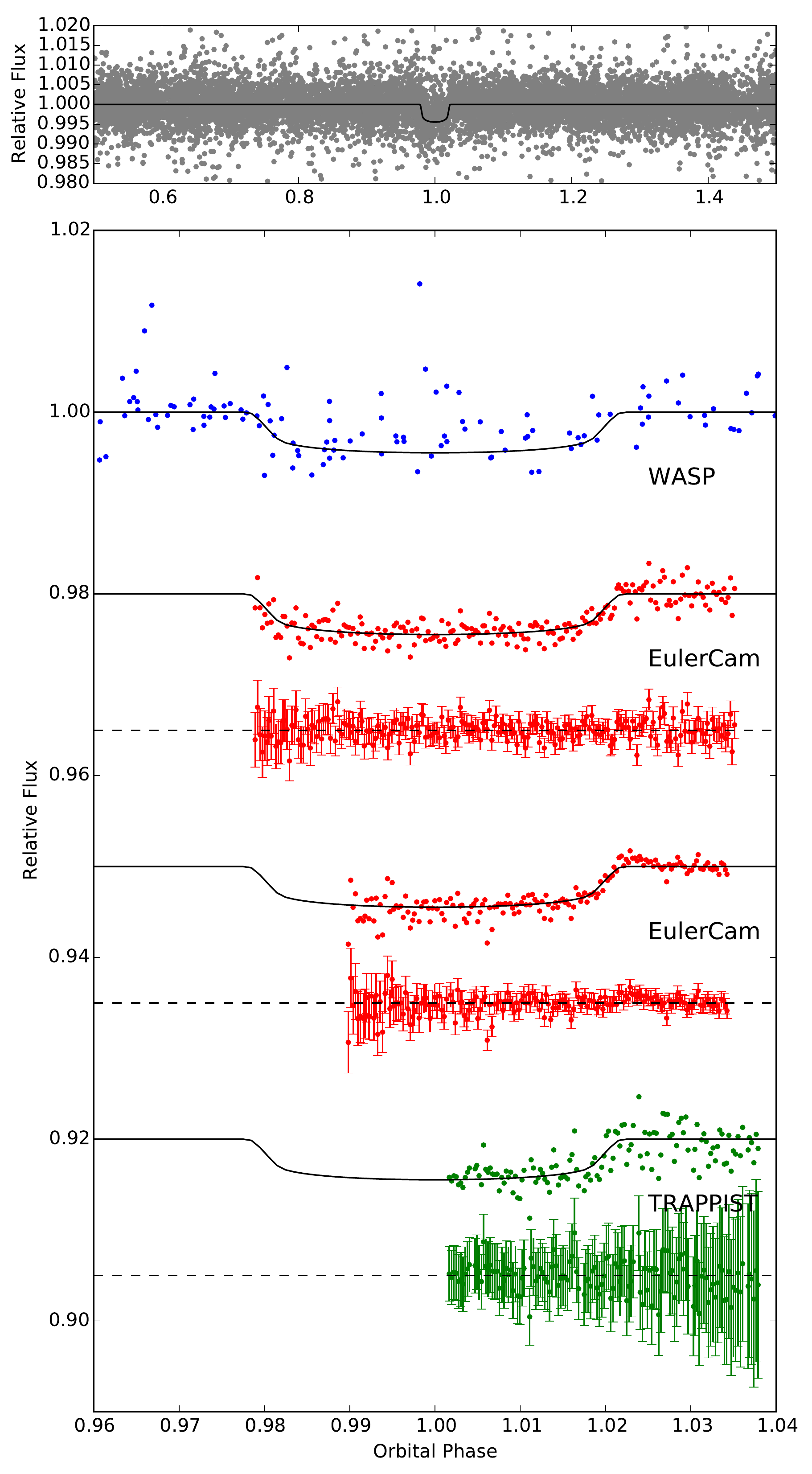}
	    \caption{\label{photomtery136} Photometry follow-up of WASP-136 observed from EULERCam (red) and TRAPPIST(green). The data are phase-folded with the ephemeris from our analysis. The light curves are assigned an arbitrary offset from the zero magnitute and are binned to a 2-minute cadence for clarity. The best-fit transit model from \citet{2002ApJ...580L.171M} is plotted as a black solid line, and the residuals of the fit are plotted directly below the light curves. }
	\end{figure}

\begin{figure}[htbp!]
	\centering
		    \includegraphics[width=0.45\textwidth]{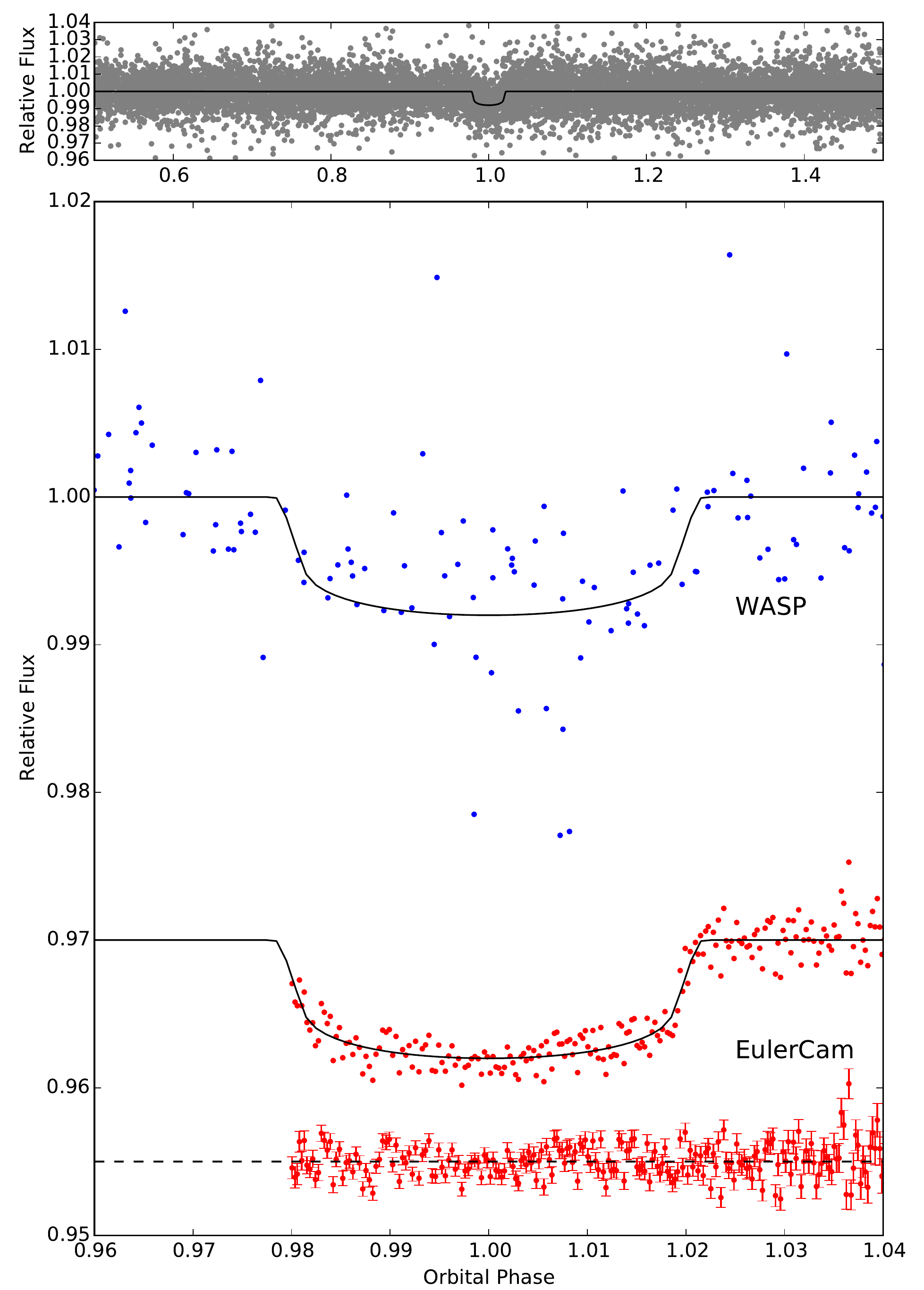}
	    \caption{\label{photomtery138}Photometry follow-up of WASP-138 observed from EULERCam (red). The data are phase-folded with the ephemeris from our analysis. The EulerCAM light curve is assigned an arbitrary offset from the zero magnitute for clarity. The best-fit transit model from \citet{2002ApJ...580L.171M} is plotted as a black solid line, and the residuals of the fit are plotted directly below the light curves.}
	\end{figure}

		\hspace{0.5cm}Multiple follow-up photometry was taken for the three stars to place better constraints on the system parameters. The photometry was obtained with EulerCam at the $1.2$ m Euler-Swiss telescopes \citep{2012A&A...544A..72L} and TRAPPIST \citep{2011Msngr.145....2J,2011EPJWC..1106002G}, which are situated at ESO La Silla Observatory in Chile, the RISE camera on the Liverpool Telescope at the Observatorio del Roque de los Muchachos on La Palma \citep{2008SPIE.7014E..6JS} and the Zeiss $1.23$ m telescope at the German-Spanish Astronomical Center at Calar Alto in Spain. The summary of our follow-up photometric observations is given in Table \ref{PhotometrySummary}, and the phase-folded light curves are shown in Figs.~\ref{photomtery127}, \ref{photomtery136} and \ref{photomtery138}, along with their best-fit transit models derived from our analysis in Sect.~\ref{mcmcanalysis}. 
		
		 \textbf{TRAPPIST:}~
		 WASP-127 and WASP-136 were both observed with the 0.6 m TRAnsiting Planets and PlanetesImals Small Telescope (TRAPPIST) robotic telescope. The telescope is equipped with a thermoelectrically cooled 2k$\times$2k CCD camera, which has a pixel scale of 0.65'' that translates into a 22'$\times$22' field of view. For details of TRAPPIST, see \citet{2011EPJWC..1106002G} and \citet{2011Msngr.145....2J}.
		 
		A partial transit of WASP-127b was observed on 2014 March 18 through a Sloan-z’ filter (effective wavelength $=896.3\pm0.8$ nm) with an exposure time of 9 seconds. The same filter was used to observe a partial transit of WASP-136b on 2014 November 24 (effective wavelength $=895.0\pm0.6$ nm), but with an exposure time of 7 seconds. Throughout the observations, the telescope was kept in focus and the positions of the stars on the detector were retained on the same few pixels, thanks to a software guiding system that regularly derives an astrometric solution for the images and sends pointing corrections to the mount when needed. The observation of WASP-136 was made towards the end of night. Therefore the increase in airmass is reflected in the uncertainties of the photometry.
		
		Data were reduced as described in \citet{2013A&A...552A..82G}. After a standard pre-reduction (bias, dark, and flat-field correction), the stellar fluxes were extracted from the images using IRAF/DAOPHOT\footnote{IRAF is distributed by the National Optical Astronomy Observatory, which is operated by the Association of Universities for Research in Astronomy, Inc., under cooperative agreement with the National Science Foundation.} \citep{1987PASP...99..191S}. For each light curve, we tested several sets of reduction parameters and chose the one giving the most precise photometry for the stars of similar brightness as the target. After a careful selection of reference stars, the transit light curves were finally obtained using differential photometry.

		 \textbf{EulerCam:}~
		  We observed one full transit of WASP-127, one partial and one full transit of WASP-136, and one full transit of WASP-138 with EulerCam \citep{2012A&A...544A..72L}. The full transit of WASP-127 was observed on 2014 April 28 with a Gunn r filter. The telescope was defocused throughout the observation with a FWHM of between 1.6 and 2.5 arcsec. A circular aperture of radius 4.7 arcsec was used along with one reference star for the extraction of the light curve.
		 
		 A partial and a full transit of WASP-136 were obtained on 2014 August 21 and 2015 August 21, respectively. Both observations were taken using a Gunn z filter and an exposure time of 50 seconds. The telescope was substantially defocused throughout both nights. The FWHM of the first night was between 1.5 and 2.3 arcsec. A circular aperture of radius 2.7 arcsec and four reference stars were used for photometry extraction. The FWHM of the second night was between 1.9 and 3.0 arcsec. We used a circular aperture of radius 4.5 arcsec along with five reference stars for the photometry reduction.
		 
		The observation of WASP-138 was carried out on 2015 December 17 with an NGTS filter (with a custom wavelength of 550 - 900 nm) and exposure times of between 50 and 85 seconds. The telescope was substantially defocused and the FWHM was between 1.3 and 2.5 arcsec. A photometric aperture of 5.6 arcsec radius was used to extract the fluxes, and one reference star was used to generate the relative light curve. See \citet{2012A&A...544A..72L} for further details on EulerCam and its data reduction procedures.
		 
		\textbf{RISE:}~
		A full transit of WASP-127 was observed with RISE \citep{2008SPIE.7014E..6JS}. The camera is equipped with a back-illuminated frame-transfer CCD of $1024\times1024$ pixels. A V+R filter (a custom filter constructed from 3 mm OG515 + 2 mm KG3, with a bandwidth of 500-900 nm) and a $2\times2$ binning of the detector were used for the observation, resulting in a pixel scale of $1.08$ arcsec/pixel. We used an exposure time of 1.5 seconds and defocused the telescope by 0.5 mm for all the observations. Images were automatically bias, dark, and flat corrected by the RISE pipeline. We selected four comparison stars for data reduction. The data were reduced with the standard IRAF apphot routines using a 1.4 pixels (4.86 arcsec) aperture. We attribute the increased scatter around mid-transit to thin clouds (see Fig.~\ref{photomtery127}).

		 \textbf{ZEISS:}~
		The Zeiss $1.23$ m telescope has a focal length of $9857.1$ mm and is equipped with the DLR-MKIII camera, which has 4k$\times$4k pixels of size 15 micron. The plate scale is $0.32$ arcsec/pixel and the field of view is $21.5\times21.5$ arcmin. It has previously been successfully used to follow-up many planetary transits (e.g. \citet{2015A&A...579A.136M}). The telescope was defocused and the exposure time was adjusted several times during the night in a range of between 65 and 105 seconds.
		 
		 The guiding camera did not operate correctly on the night of our observations, and the usual precision was not achieved. The CCD was windowed to decrease the readout time and therefore sped up the cadence of the observations. The night was not photometric, and several clouds disturbed the observations. The data were reduced using a revised version of the \textsc{defot} code \citep{2014MNRAS.444..776S}. In brief, the scientific images were calibrated and the photometry was extracted by the standard aperture-photometry technique. The resulting light curve was normalised to zero magnitude by fitting a straight line to the out-of-transit data.

\section{\label{results}Results}
		\subsection{\label{SpectralAnalysis}Stellar parameters}
%		stellar spectral analysis
		\hspace{0.5cm}The CORALIE spectra of the individual host stars were co-added to produce spectra for analysis using the methods described in \citet{2013MNRAS.428.3164D}. The H$\alpha$ line was used to estimate the effective temperature ($\rm T_{eff}$), and the Na~{\sc i} D and Mg~{\sc i} b lines were used as diagnostics of the surface gravity ($\log g$). The iron abundances were determined from equivalent-width measurements of several clean and unblended Fe~{\sc i} lines and are given relative to the solar value presented in \citet{2009ARA&A..47..481A}. The quoted abundance errors include that given by the uncertainties in $\rm T_{eff}$ and $\log g$, in addition to the scatter due to measurement and atomic data uncertainties. The projected rotation velocities ($v \sin i$) were determined by fitting the profiles of the Fe~{\sc i} lines after convolving with the CORALIE instrumental resolution ($R$ = 55\,000) and macroturbulent velocities adopted from the calibration of \citet{2014MNRAS.444.3592D}. The result of our spectral analysis is listed in Table \ref{StellarParam}.

%		look up star catalogues and see if these stars were previously identified.

\begin{table}
\centering
\small
		\caption{\label{StellarParam}Stellar parameters of WASP-127, WASP-136, and WASP-138 obtained from spectral analysis.}
\begin{tabular}{c|c|c|c}
\hline
Parameter & WASP-127 & WASP-136 & WASP-138 \\ 
\hline 
\hline
$\rm T_{eff}$ (K) & $5750 \pm 100$ & $6250 \pm 100$ & $6300 \pm 100$ \\ 
%\hline 
log g & $3.9 \pm 0.1$ & $3.9 \pm 0.1$ & $4.1 \pm 0.1$ \\ 
%\hline 
${v \sin i}$ (km $\rm s^{-1}$) & $0.3 \pm 0.2$ & $13.1 \pm 0.8$ & $7.7 \pm 1.1$ \\ 
%\hline 
$\rm [Fe/H]$ & $-0.18 \pm 0.06$ & $-0.18 \pm 0.10$ & $-0.09 \pm 0.10$ \\ 
%\hline 
log A(Li) & $1.97 \pm 0.09$ & $2.50 \pm 0.08$ & $2.20 \pm 0.08$ \\ 
%\hline 
Mass (${\rm M}_{\odot}$) & $1.31 \pm 0.05$ & $1.38 \pm 0.08$ & $1.20 \pm 0.03$ \\ 
%\hline 
Radius (${\rm R}_{\odot}$) & $1.33 \pm 0.03$ & $2.07 \pm 0.24$ & $1.43 \pm 0.02$ \\ 
%\hline 
Sp. Type & G5 & F5 & F9 \\ 
%\hline 
Distance (pc) & $102 \pm 12 $ & $164 \pm 18$ & $308 \pm 51$ \\ 
\hline 
\end{tabular} 	
\end{table}

\begin{figure}[htbp!]
	    \includegraphics[width=0.5\textwidth]{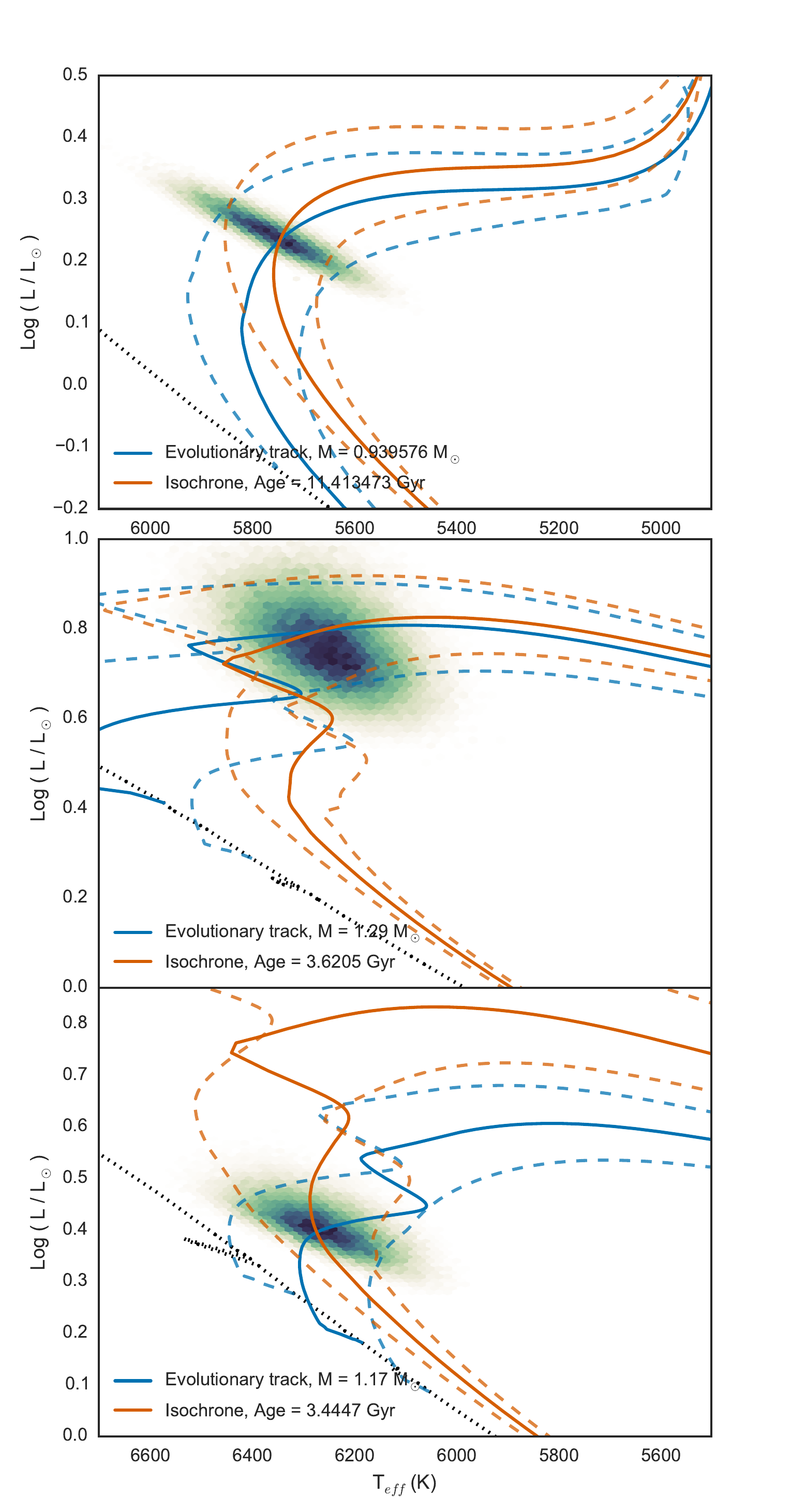}
	    \caption{\label{StellarTracks}Stellar mass and age analysis of WASP-127(upper panel), WASP-136(middle panel) and WASP-138(lower panel) using \texttt{BAGEMASS}. The dotted black line is the ZAMS. The solid blue line is the mass evolutionary track, and the blue dashed tracks on either side are for the 1$\sigma$ error of the mass. The solid orange line is the stellar age isochrone and the orange dashed lines represents the 1$\sigma$ error. The density of MCMC samples is shown in the colour scale of the plotted posterior distribution. }
	\end{figure}

We used the open-source \texttt{BAGEMASS}\footnote{https://sourceforge.net/projects/bagemass/} code \citep{2015A&A...575A..36M} to estimate the masses and ages of the three stars. The stellar mass and age were derived by calculating the stellar model grid of a single star using the \texttt{GARSTEC} stellar evolution code \citep{2008Ap&SS.316...99W}. A Bayesian method was then applied to sample the probability distribution of the posterior mass and age. The result of our stellar mass and age analysis is shown in Table \ref{StellarAge}. The probability distribution along with the best-fit stellar evolutionary tracks and isochrones are plotted in Fig.~\ref{StellarTracks}. WASP-136 is an F-type star with an age estimated to be $3.62 \pm 0.70$ Gyr. It also has a surface gravity of log g $= 3.9 \pm 0.1$. This implies that WASP-136 is a subgiant star that is evolving off the main sequence.
We also applied the relation from \citet{2007ApJ...669.1167B} to derive the gyrochronological ages ($\tau_{\rm gyro}$) of WASP-136 and WASP-138. From the stellar rotation period calculated from the measured $v \sin i$ and radius, the estimated $\tau_{\rm gyro}$ of WASP-136 and WASP-138 are $1.3_{-0.6}^{+1.2}$ Gyr and $2.7_{-1.3}^{+2.5}$ Gyr, respectively. The $v \sin i$ value of WASP-127 is too low for a sensible estimate. The $v \sin i$ values give upper limits on the rotation periods, hence the $\tau_{\rm gyro}$ can only provide a lower limit here. There are discrepancies between the isochronal ages and $\tau_{\rm gyro}$ of WASP-136 and WASP-138, which may have arisen as a result of tidal interactions. In gyrochronology, the rate of stellar surface rotation is used to determine the stellar age. However, currently available gyrochronology models do not predict the observed rotation rate of older stars. Evolved stars exist whose rotation rate is spun up by the tidal forces of the planets, which causes $\tau_{\rm gyro}$ to appear significantly lower than the isochronal ages \citep{2015A&A...577A..90M,2016Natur.529..181V}. This suggests that $\tau_{\rm gyro}$ is a less suitable way of estimating the stellar age. We searched for rotational modulation of the WASP photometry using the method of \citet{2011PASP..123..547M}. No rotational modulation was found above 2 mmag, suggesting that the host stars are inactive.

\begin{table}
\centering
		\caption{\label{StellarAge}Stellar mass and age estimates of WASP-127, WASP-136, and WASP-138 from \texttt{BAGEMASS}. The isochronal ages are presented in the $\tau_{\rm iso}$ column and the gyrochronological ages are presented in the $\tau_{\rm gyro}$ column. }
\begin{tabular}{c c c c}
\hline
Star & Mass [$M_{\odot}$] & $\tau_{\rm iso}$ [Gyr] & $\tau_{\rm gyro}$ [Gyr]\\ 
\hline 
\hline
WASP-127 & $0.93 \pm 0.04$ & $11.41 \pm 1.80$ & $v \sin i$ too low\\ 
%\hline 
WASP-136 & $1.29 \pm 0.08$& $3.62 \pm 0.70$ & $>1.3_{-0.6}^{+1.2}$\\ 
%\hline 
WASP-138 & $1.17 \pm 0.06$ & $3.44 \pm 0.93$ & $>2.7_{-1.3}^{+2.5}$\\ 
\hline 
\end{tabular} 	
\end{table}	

		\subsection{\label{mcmcanalysis}System parameters from MCMC analysis}
		
		\hspace{0.5cm}The Markov chain Monte Carlo (MCMC) method was used to derive the system parameters. We simultaneously analysed the WASP photometry, the follow-up photometry, and CORALIE and SOPHIE RV measurements. The details of our method are described in \citet{2007MNRAS.380.1230C} and \citet{2008MNRAS.385.1576P}. In summary, the transit light curves were modelled with analytical functions from \citet{2002ApJ...580L.171M}, and the stellar limb-darkening was accounted for using the non-linear four-component model of \citet{2000A&A...363.1081C,2004A&A...428.1001C}. We also applied a linear decorrelation to remove systematic trends in our photometry. The parameters used in our MCMC analysis are the mid-transit epoch $\rm T_0$, the period $\rm P$, the planet-to-stellar-size ratio (a proxy for the transit depth) $\Delta F$, the transit duration $\rm T_{14}$, the impact parameter $b$, the stellar metallicity $\rm [Fe/H]$, the stellar effective temperature $\rm T_{eff}$, the stellar reflex velocity $K_1$, and the Lagrangian elements $\sqrt{e}\cos(\omega)$ and $\sqrt{e}\sin(\omega)$ (where $e$ is the eccentricity and $\omega$ is the longitude of periastron). These values were randomly perturbed at each step of the burn-in phase and the minimum $\chi ^{2}$ of the model was calculated. The Metropolis-Hastings method \citep{1953JChPh..21.1087M,Hastings70} was applied to derive the best-fit solution to the set of parameters, and the median of the posterior distribution was used as the final solution along with the 1$\sigma$ uncertainties. The result of our MCMC analysis is presented in Table \ref{SystemParam} and the best-fit RV curves and the best-fit transit light curves are presented in Figs.~\ref{RV127_wosophie}-\ref{RV138} and Figs.~\ref{photomtery127}-\ref{photomtery138}, respectively.
		
		\textbf{WASP-127:}~
		We imposed a main-sequence mass-radius constraint in the MCMC analysis of WASP-127. However, the constraint offers a solution where the posterior stellar parameters do not agree with our spectral analysis as described in Sect.~\ref{SpectralAnalysis}. We therefore relaxed the main-sequence constraint for our analysis. The solution does not give convincing evidence of an eccentric orbit ($\chi_{circ}^{2} = 37.2$ and $\chi_{ecc}^{2} = 37.3$), and a circular orbit is adopted. 
		
		Both CORALIE and SOPHIE RV measurements show some dispersion from the fit. However, CORALIE RVs suggest an in-phase variation. Initial analysis with both sets of data suggests a planet with a much lower mass that is unlikely to be detected by both spectrographs. We therefore increased the size of the error bars of the SOPHIE RVs by a multiplication factor of 2, to cause the reduced chi-square statistics ($\chi_{\rm reduced}^2$) to converge to unity. The solution using the original weighting of the SOPHIE RVs has a $\chi_{\rm reduced}^2$ value of $1.31$. When the weighting of the SOPHIE RVs is decreased, the $\chi_{\rm reduced}^2$ becomes $0.91$. We decided to look for a best-fit solution with the latter case because the fit was improved.
		
		We explored the solutions with and without the set of SOPHIE RV measurements. For the case using only the CORALIE RVs, the best-fit reflex radial velocity value is $21.4 \pm 2.8$ ms$^{-1}$. The solution using both the CORALIE and SOPHIE RVs gives a best-fit reflex radial velocity of $18.7 \pm 2.7$ ms$^{-1}$. The two solutions give a 1$\sigma$ agreement in the reflex radial velocity, and Table \ref{W127SystemParam} shows the solutions from both analyses. We attribute the dispersion of the residuals to the very low mass of the planet, which is challenging for the spectrographs to detect. Higher precision follow-up RV measurements is recommended in the future to accurately determine the mass of the planet.

		\textbf{WASP-136:}~
		As expected from a subgiant star, a main-sequence mass-radius constraint gives an unrealistic solution for the stellar metallicity and the effective temperature. We used the $\chi^2$ statistics to test for the goodness of fit of our model. There is no evidence supporting an eccentric orbit ($\chi_{circ}^2 = 46.6$ and $\chi_{ecc}^2 = 43.5$), hence we adopt the circular orbit and relaxed the main-sequence constraint.

		\textbf{WASP-138:}~
		Using all the follow-up photometry and RV measurements, we find that imposing a main-sequence constraint has an insignificant effect on the solution. There is no evidence suggesting an eccentric orbit ($\chi_{circ}^2=11.2$ and $\chi_{ecc}^2=10.4$). We therefore adopt the solution with a circular orbit with no main-sequence constraint.

		\begin{table}[!t]
		\caption{\label{W127SystemParam}WASP-127 parameters from the MCMC analysis. The solutions are derived with and without SOPHIE RV data to test for the effect of the RV dispersion. The resulting RV fits show a 1$\sigma$ agreement between the two solutions.}
		\tiny
%		\centering
		\begin{tabular}{ c c c }
		\hline 
		Parameter (Unit) & Solution without SOPHIE& Solution with SOPHIE \\ 
		\hline 
		P (d) & $4.178062 \pm 0.000002$ & $4.178062 \pm 0.000002$\\ 
%		\hline 
		T$_0$ (BJD) & $2457248.74131 \pm 0.000160$ & $2457248.74126 \pm 0.000184$  \\ 
%		\hline 
		$\Delta F = R_{pl}^2 / R_{*}^2$ & $0.01037 \pm 0.00015$ & $0.01050 \pm 0.00017$ \\ 
%		\hline 
		$\rm T_{14}$ (d) & $0.1795 \pm 0.0007$ & $0.1802 \pm 0.0009$\\ 
%		\hline 
		b & $0.15_{-0.11}^{+0.09} $ & $0.25_{-0.16}^{+0.08} $  \\ 
%		\hline 
		i ($ {}^{\circ}$) & $88.7_{-0.6}^{+0.8}$ & $88.1_{-0.7}^{+1.2}$   \\ 
%		\hline
		$\rm M_* (M_{\odot})$ & $1.08 \pm 0.03$ & $1.08 \pm 0.03$  \\
%		\hline
		$\rm R_* (R_{\odot})$ & $1.39 \pm 0.03$ & $1.42 \pm 0.05$  \\
%		\hline
		$\rm log\, g_*$ (cgs) & $4.18 \pm 0.01$ & $4.17 \pm 0.0.02$ \\
%		\hline
		$\rho_* (\rho_{\odot})$ & $0.404 \pm 0.015$ & $0.380 \pm 0.031$ \\
%		\hline
		$\rm T_{eff}$ (K) & $5620 \pm 85$ & $5639 \pm 90$  \\
%		\hline
		${\rm M}_{pl} (M_{J})$ & $0.18 \pm 0.02$ & $0.16 \pm 0.02$ \\ 
%		\hline 
		${\rm R}_{pl} (R_{J})$ & $1.37 \pm 0.04$ & $1.41 \pm 0.06$ \\ 
%		\hline
		$\rm log\,g_{pl}$ (cgs) & $2.33 \pm 0.06$ & $2.25 \pm 0.7$ \\ 
%		\hline
		$\rho_{pl} (\rho_{J})$ & $0.068_{-0.010}^{+0.010}$ & $0.055_{-0.009}^{+0.011}$  \\
%		\hline 
		a (au) & $0.0520 \pm 0.0005$ & $0.0522 \pm 0.0005$ \\ 
		
		$\rm T_{pl,A=0}$ (K) &$1400 \pm 24$ & $1417 \pm 32$ \\
		\hline
		\end{tabular} 
		\end{table}

		\begin{table}[!t]
		\caption{\label{SystemParam}System parameters of WASP-136 and WASP-138 from the MCMC analysis.}
		\tiny
		\begin{tabular}{ c c c }
		\hline 
		Parameter (Unit) & WASP-136b & WASP-138b \\ 
		\hline 
		P (d) &  $5.215357 \pm 0.000006$ & $3.634433 \pm 0.000005$ \\ 
%		\hline 
		T$_0$ (BJD) & $2456776.90615 \pm 0.00109$ & $2457326.62183 \pm 0.000319$ \\ 
%		\hline 
		$\Delta F = R_{pl}^2 / R_{*}^2$ & $0.00411 \pm 0.00015$ & $0.00683 \pm 0.00013$ \\ 
%		\hline 
		$\rm T_{14}$ (d) & $0.2272 \pm 0.0033$ & $0.1572 \pm 0.0012$ \\ 
%		\hline 
		b & $0.59_{-0.14}^{+0.08} $ & $0.19_{0.15}^{0.12}$ \\ 
%		\hline 
		i ($ {}^{\circ}$) & $84.7_{-1.3}^{+1.6}$  & $88.5_{-1.2}^{+0.9}$ \\ 
%		\hline
		$\rm M_* (M_{\odot})$ & $1.41 \pm 0.07$ & $1.22 \pm 0.05$ \\
%		\hline
		$\rm R_* (R_{\odot})$ & $2.21 \pm 0.22$ & $1.36 \pm 0.05$ \\
%		\hline
		$\rm log\, g_*$ (cgs) & $3.90 \pm 0.06$ & $4.25 \pm 0.02$ \\
%		\hline
		$\rho_* (\rho_{\odot})$ & $0.132 \pm 0.030$ & $0.488 \pm 0.044$ \\
%		\hline
		$\rm T_{eff}$ (K) & $6260 \pm 100$ & $6272 \pm 96$ \\
%		\hline
		${\rm M}_{pl} (M_{J})$ & $1.51 \pm 0.08$ & $1.22 \pm 0.08$ \\ 
%		\hline 
		${\rm R}_{pl} (R_{J})$ & $1.38 \pm 0.16$ & $1.09 \pm 0.05$ \\ 
%		\hline
		$\rm log\, g_{pl}$ (cgs) & $3.26 \pm 0.09$ & $3.36 \pm 0.04$ \\ 
%		\hline
		$\rho_{pl} (\rho_{J})$ & $0.581_{-0.148}^{+0.230}$ & $0.92_{-0.146}^{+0.097}$ \\
%		\hline 
		a (au) & $0.0661 \pm 0.0012$& $0.0494 \pm 0.0007$ \\ 
		
		$\rm T_{pl,A=0}$ (K) &$1742 \pm 82$ & $1590 \pm 31$ \\
		\hline
		\end{tabular} 
		\end{table}

		\section{\label{discussion}Discussion and conclusion}
		
		\subsection{WASP-127b}
		\hspace{0.5cm}From our best-fit MCMC solution, we obtain a planet with a mass of $0.18 \pm 0.02 ~\rm M_{J}$ and a radius of $1.37 \pm 0.04 ~\rm R_{J}$ ($\rm M_{pl} = 0.16 \pm 0.02 ~M_J$ and $\rm R_{pl} = 1.41 \pm 0.06 ~R_J$ for the case where RVs from both CORALIE and SOPHIE were included for analysis). This means that WASP-127b has a density of $0.07_{-0.01}^{+0.01} ~\rho_{J}$, making it one of the lowest density planets ever discovered. It is also the planet with the second lowest mass discovered by WASP, only more massive than WASP-139b \citep{2016arXiv160404195H}. 
		
		Compared to the standard coreless model of \citet{2007ApJ...659.1661F}, WASP-127b is over $30\%$ larger than expected for a planet with an orbit at 0.045 au around a 4.5 Gyr solar-type star. From Sect.~\ref{SpectralAnalysis}, WASP-127 is estimated to be much older than the Sun. Hence the theoretical radius of the planet should be even smaller. The anomalously large radius of WASP-127b could be explained by several inflation mechanisms. One such mechanism is tidal heating \citep{2001ApJ...548..466B,2003ApJ...592..555B}, where the planetary interior receives heat energy as the orbit circularises. As with many short-period gas giants, the orbit of WASP-127b may have shrunk and migrated to its current position through planet-planet scattering \citep{2008ApJ...686..621F} or Kozai mechanism \citep{2007ApJ...669.1298F,1962AJ.....67..591K,1962P&SS....9..719L}. The planetary orbit may also have been tidally circularised during the migration process, which results in rapid transfer of energy to its interior, inflating the radius of the planet. This energy transfer process occurs rapidly at the early stages of the evolution history of the system, hence it is unclear how efficient this way of heating up the planetary interior is for such an old system. On the other hand, the atmosphere of WASP-127b could have an enhanced opacity resulting from enhanced metallicity \citep{2007ApJ...661..502B}. This can delay the cooling effect and maintain the inflated radius for a longer time. 
		\citet{2011ApJ...738....1B} suggested that the Ohmic heating mechanism could account for the large planetary radii. The interaction between the planetary magnetic field and the flow of ionised atmospheric heavy elements could induce an electro-motive force. This reaction can drive electrical currents throughout the planet and lead to inflation as the planet heats up. The Ohmic dissipation, however, is limited by the depth of the dissipation. The \citet{2012ApJ...757...47H} model suggests that the convective zone boundary will move deeper if Ohmic dissipation occurs outside the convection zone, which in turn will slow down the efficiency of the planet cooling process. Recently, \citet{2016ApJ...818....4L} have suggested that a planet whose radius was inflated through internal heating could re-inflate again as its host star moves towards the RGB phase because of increased irradiation. Their studies also showed that re-inflation of a planet is most likely for low-mass and short-period planets. WASP-127 is estimated to have a main-sequence lifetime of approximately $\rm t_{MS} = t_{\odot}(M/M_{\odot})^{-2.5} \approx 8$ Gyr, where $\rm t_{\odot}$ is the solar main-sequence lifetime, $\rm M_odot$ is the solar mass and M is the stellar mass. WASP-127 has an isochronal age of $11.41 \pm 1.80$ Gyr, implying that the star is at the end of the main-sequence phase. Therefore,  WASP-127b may be undergoing a phase of re-inflation as its host star enters the subgiant branch.
		
\begin{figure}[htbp!]
	    \includegraphics[width=0.5\textwidth]{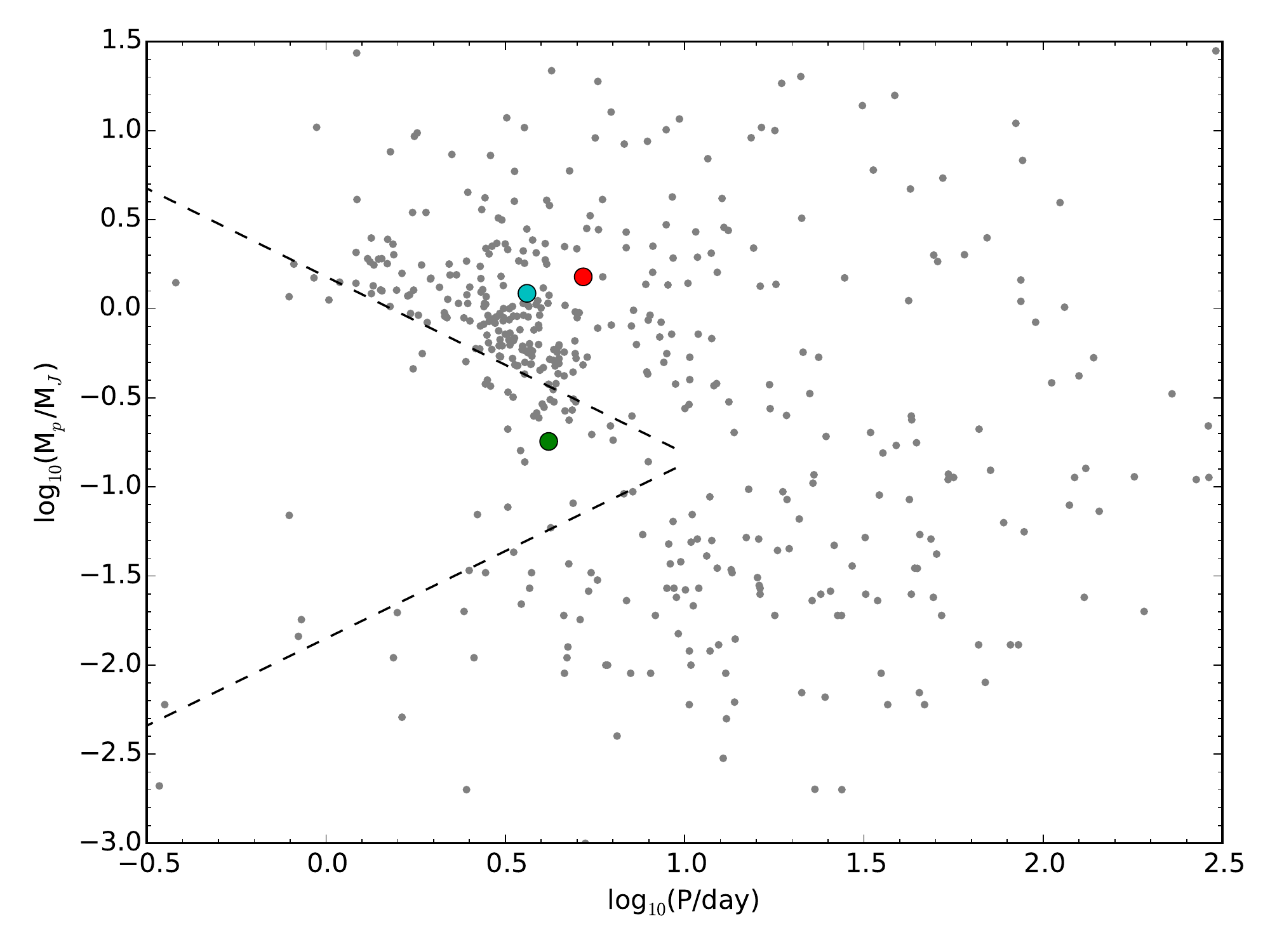}
	    \caption{\label{NeptuneDesert} Planetary mass against orbital period plot. Exoplanet data are taken from the NASA Exoplanet Archive (\protect\url{http://exoplanetarchive.ipac.caltech.edu/}) and are plotted as grey dots. WASP-127b, WASP-136b, and WASP-138b are represented by green, red, and cyan circles, respectively. The black dashed lines are the upper and lower boundaries of the Neptune desert as defined by Mazeh et al. 2016. WASP-127b falls in this Neptune desert between Jovian and super-Earth planets.}
	\end{figure}

		WASP-127b is a highly inflated planet that shares many similarities with a number of low-density planets such as WASP-39b \citep{2011A&A...531A..40F}, WASP-113b \citep{2016arXiv160702341B}, HAT-P-8b \citep{2015AJ....150...49B}, HAT-P-11b \citep{2010ApJ...710.1724B}, HAT-P-47b, and HAT-P-48b \citep{2016arXiv160604556B}. Assuming it has an atmosphere identical to that of Jupiter ($\mu = 2.2$u, where the atomic mass unit is u $=1.66\times10^{-27}$ kg), its scale height would be $\rm H \approx 2500 \pm 400 ~km$. WASP-127b also orbits a bright G-type star of $\rm V_{mag} = 10.172$, making it an excellent candidate for transmission spectroscopy. Figure \ref{NeptuneDesert} shows the planetary mass as a function of the orbital period. WASP-127b falls into the short-period Neptune desert \citep{2016A&A...589A..75M}, a region between Jovian and super-Earth planets with a lack of detected planets. The presence of such a region indicates the existence of two unique types of short-period planets, hot Jupiters and super-Earths, which have very different formation and evolution mechanisms. It also suggests that short-period planets with intermediate masses are likely to be destroyed.

		\subsection{WASP-136b}
		\hspace{0.5cm}WASP-136b is an inflated hot-Jupiter transiting a bright F5 host star ($\rm V_{mag} = 9.928$). The best-fit MCMC solution of the system gives a planetary mass of $1.51 \pm 0.08 ~\rm M_{J}$ and a radius of $1.38 \pm 0.16 ~\rm R_{J}$ which yields a planet density of $0.58_{-0.15}^{+0.23} ~\rho_{J}$. The estimated main-sequence lifetime of WASP-136 is approximately $\rm t_{MS} \approx 4$ Gyr. In Sect.~\ref{SpectralAnalysis} we estimated the isochronal age of WASP-136 to be $\sim 3.6 \pm 0.7$ Gyr. This suggests that WASP-136 is at the end of the main-sequence phase. The density and surface gravity of WASP-136 implies that the star is a subgiant, which is consistent with the age estimate. The radius of WASP-136b is approximately $25\%$ larger than predicted from the planet evolution model of \citet{2007ApJ...659.1661F} for a coreless planet. Like many bloated hot-Jupiters (WASP-54b: \citealt{2013A&A...551A..73F}; WASP-78b and WASP-79b: \citealt{2012A&A...547A..61S}; WASP-142b: \citealt{2016arXiv160404195H}), WASP-136b receives a stronger irradiation from its F-type host star compared to a G-type, which can lead to a more inflated planet. Similar to EPIC 211351816.01 \citep{2016arXiv160605818G}, the inflation mechanism of WASP-136b could follow the class I model from \citet{2016ApJ...818....4L}, where WASP-136b might be heated up through the deposit of some of the incident stellar irradiation into the planetary interior. Consequently, when the host of WASP-136b enters the subgiant branch, the planet could receive an increased stellar irradiation that will lead to re-inflation of the planet radius.
		
				The existence of short-period hot Jupiters around subgiant stars is rare. The lack of giant planets at short orbital distances may be a consequence of tidal disruption \citep{2013ApJ...772..143S}. A planet residing inside the synchronous orbit could experience strong tidal forces and spiral inwards towards the star. When the planet eventually reaches the Roche limit, the tidal force becomes stronger than the planet's gravitational force, and the planet will be tidally disrupted. WASP-136b has an orbital distance of $0.0661$ au and is approximately 46 times the Roche limit. We followed the derivation of \citet{2010ApJ...725.1995M} and estimated that WASP-136b has a remaining lifetime of $\sim 0.683$ Gyr. Furthermore, during the post-main-sequence phase, the stellar radius expands and will eventually lead to the engulfment of the planet \citep{2009ApJ...705L..81V}.

		\subsection{WASP-138b}
		\hspace{0.5cm}Our best-fit solution gives a system with a planetary mass and radius of $1.22 \pm 0.08 ~\rm M_{J}$ and $1.09 \pm 0.05 ~\rm R_{J}$. The density of WASP-138b is $0.92_{-0.15}^{+0.10} ~\rm \rho_J$. It orbits around an F9 star with a metallicity of $\rm [Fe/H]=-0.09 ~dex$, slightly more metal-poor than the Sun. WASP-138b is a hot Jupiter that has similar characteristics to many short-period gas giants with a period of $\sim 3$days, for example, WASP-35b \citep{2011AJ....142...86E} and WASP-141b \citep{2016arXiv160404195H}. With an isochronal age of $3.4 \pm 0.9$ Gyr, the planet evolution model of \citet{2007ApJ...659.1661F} suggests that WASP-138b has a core mass of $\sim 10 ~\rm M_\oplus$ of heavy elements.

\hspace{0.5cm}Many interesting large short-period exoplanets are being discovered by ground-based surveys. Discoveries such as that of WASP-127b will provide exceptional targets for future missions such as JWST and for further characterisation. With the diverse range of exoplanets, it is important to understand the mass-radius relation and distinguish the mechanisms responsible for each distinct class of planet. This will ultimately help our understanding of the dynamics of planetary systems.

\begin{acknowledgements}
The SuperWASP Consortium consists of astronomers primarily from Warwick, Queens University Belfast, St Andrews, Keele, Leicester, The Open University, Isaac Newton Group La Palma and Instituto de Astrofsica de Canarias. The SuperWASP-N camera is hosted by the Issac Newton Group on La Palma and WASPSouth is hosted by SAAO. We are grateful for their support and assistance. Funding for WASP comes from consortium universities and from the UK’s Science and Technology Facilities Council. Based on observations made with SOPHIE spectrograph mounted on the 1.9-m telescope at Observatoire de Haute-Provence (CNRS), France and at the ESO La Silla Observatory (Chile) with the CORALIE Echelle spectrograph mounted on the Swiss telescope. TRAPPIST is funded by the Belgian Fund for Scientific Research (Fonds National de la Recherche Scientifique, FNRS) under the grant FRFC 2.5.594.09.F, with the participation of the Swiss National Science Fundation (SNF). M.~Gillon and E.~Jehin are FNRS Research Associates. L. Delrez acknowledges support of the F.R.I.A. fund of the FNRS. The Liverpool Telescope is operated on the island of La Palma by Liverpool John Moores University in the Spanish Observatorio del Roque de los Muchachos of the Instituto de Astrofisica de Canarias with financial support from the UK Science and Technology Facilities Council. The research leading to these results has received funding from the European Community’s Seventh Framework Programme (FP7/2007-2013)under grant agreement number RG226604 (OPTICON). D.J.A., D.P., and A.S.B. acknowledge funding from the European Union Seventh Framework programme (FP7/2007-2013) under grant agreement No. 313014 (ETAEARTH). DJAB acknowledges support from the UKSA and the University of Warwick. SCCB acknowledges support by the Fundação para a Ciência e a Tecnologia (FCT) through the Investigador FCT Contract No. IF/01312/2014 and the grant reference PTDC/FIS- AST/1526/2014.

\end{acknowledgements}

\bibliographystyle{aa}
\bibliography{wasp127}{}

\begin{thebibliography}{78}
\expandafter\ifx\csname natexlab\endcsname\relax\def\natexlab#1{#1}\fi

\bibitem[{{Alsubai} {et~al.}(2013){Alsubai}, {Parley}, {Bramich}, {Horne},
  {Collier Cameron}, {West}, {Sorensen}, {Pollacco}, {Smith}, \&
  {Fors}}]{2013AcA....63..465A}
{Alsubai}, K.~A., {Parley}, N.~R., {Bramich}, D.~M., {et~al.} 2013, \actaa, 63,
  465

\bibitem[{{Anderson} {et~al.}(2010){Anderson}, {Hellier}, {Gillon}, {Triaud},
  {Smalley}, {Hebb}, {Collier Cameron}, {Maxted}, {Queloz}, {West}, {Bentley},
  {Enoch}, {Horne}, {Lister}, {Mayor}, {Parley}, {Pepe}, {Pollacco},
  {S{\'e}gransan}, {Udry}, \& {Wilson}}]{2010ApJ...709..159A}
{Anderson}, D.~R., {Hellier}, C., {Gillon}, M., {et~al.} 2010, \apj, 709, 159

\bibitem[{{Asplund} {et~al.}(2009){Asplund}, {Grevesse}, \&
  {Sauval}}]{2009ARA&A..47..481A}
{Asplund}, M., {Grevesse}, N., \& {Sauval}, A.~J., e.~a. 2009, \araa, 47, 481

\bibitem[{{Bakos} {et~al.}(2013){Bakos}, {Csubry}, {Penev}, {Bayliss},
  {Jord{\'a}n}, {Afonso}, {Hartman}, {Henning}, {Kov{\'a}cs}, {Noyes},
  {B{\'e}ky}, {Suc}, {Cs{\'a}k}, {Rabus}, {L{\'a}z{\'a}r}, {Papp}, {S{\'a}ri},
  {Conroy}, {Zhou}, {Sackett}, {Schmidt}, {Mancini}, {Sasselov}, \&
  {Ueltzhoeffer}}]{2013PASP..125..154B}
{Bakos}, G.~{\'A}., {Csubry}, Z., {Penev}, K., {et~al.} 2013, \pasp, 125, 154

\bibitem[{{Bakos} {et~al.}(2016){Bakos}, {Hartman}, \&
  {Torres}}]{2016arXiv160604556B}
{Bakos}, G.~{\'A}., {Hartman}, J.~D., \& {Torres}, G., e.~a. 2016, ArXiv
  e-prints [\eprint[arXiv]{1606.04556}]

\bibitem[{{Bakos} {et~al.}(2002){Bakos}, {L{\'a}z{\'a}r}, \&
  {Papp}}]{2002PASP..114..974B}
{Bakos}, G.~{\'A}., {L{\'a}z{\'a}r}, J., \& {Papp}, I., e.~a. 2002, \pasp, 114,
  974

\bibitem[{{Bakos} {et~al.}(2010){Bakos}, {Torres}, \&
  {P{\'a}l}}]{2010ApJ...710.1724B}
{Bakos}, G.~{\'A}., {Torres}, G., \& {P{\'a}l}, A., e.~a. 2010, \apj, 710, 1724

\bibitem[{{Baranne} {et~al.}(1996){Baranne}, {Queloz}, {Mayor}, {Adrianzyk},
  {Knispel}, {Kohler}, {Lacroix}, {Meunier}, {Rimbaud}, \&
  {Vin}}]{1996A&AS..119..373B}
{Baranne}, A., {Queloz}, D., {Mayor}, M., {et~al.} 1996, \aaps, 119, 373

\bibitem[{{Barnes}(2007)}]{2007ApJ...669.1167B}
{Barnes}, S.~A. 2007, \apj, 669, 1167

\bibitem[{{Barros} {et~al.}(2016){Barros}, {.~Brown}, {H{\'e}brard}, {G{\'o}mez
  Maqueo Chew}, {Anderson}, {Boumis}, {Delrez}, {Hay}, {Lam}, {Llama}, {Lendl},
  {McCormac}, {Skiff}, {Smalley}, {Turner}, {Vanhuysse}, {Armstrong}, {Boisse},
  {Bouchy}, {Collier Cameron}, {Faedi}, {Gillon}, {Hellier}, {Jehin}, {Liakos},
  {Meaburn}, {Osborn}, {Pepe}, {Plauchu-Frayn}, {Pollacco}, {Queloz}, {Rey},
  {Spake}, {S{\'e}gransan}, {Triaud}, {Udry}, {Walker}, {Watson}, {West}, \&
  {Wheatley}}]{2016arXiv160702341B}
{Barros}, S.~C.~C., {.~Brown}, D.~J.~A., {H{\'e}brard}, G., {et~al.} 2016,
  ArXiv e-prints [\eprint[arXiv]{1607.02341}]

\bibitem[{{Batygin} {et~al.}(2011){Batygin}, {Stevenson}, \&
  {Bodenheimer}}]{2011ApJ...738....1B}
{Batygin}, K., {Stevenson}, D.~J., \& {Bodenheimer}, P.~H. 2011, \apj, 738, 1

\bibitem[{{Bayliss} {et~al.}(2015){Bayliss}, {Hartman}, \&
  {Bakos}}]{2015AJ....150...49B}
{Bayliss}, D., {Hartman}, J.~D., \& {Bakos}, G.~{\'A}., e.~a. 2015, \aj, 150,
  49

\bibitem[{{Bodenheimer} {et~al.}(2003){Bodenheimer}, {Laughlin}, \&
  {Lin}}]{2003ApJ...592..555B}
{Bodenheimer}, P., {Laughlin}, G., \& {Lin}, D.~N.~C. 2003, \apj, 592, 555

\bibitem[{{Bodenheimer} {et~al.}(2001){Bodenheimer}, {Lin}, \&
  {Mardling}}]{2001ApJ...548..466B}
{Bodenheimer}, P., {Lin}, D.~N.~C., \& {Mardling}, R.~A. 2001, \apj, 548, 466

\bibitem[{{Borucki} {et~al.}(2010){Borucki}, {Koch}, {Basri}, {Batalha},
  {Brown}, {Caldwell}, {Caldwell}, {Christensen-Dalsgaard}, {Cochran},
  {DeVore}, {Dunham}, {Dupree}, {Gautier}, {Geary}, {Gilliland}, {Gould},
  {Howell}, {Jenkins}, {Kondo}, {Latham}, {Marcy}, {Meibom}, {Kjeldsen},
  {Lissauer}, {Monet}, {Morrison}, {Sasselov}, {Tarter}, {Boss}, {Brownlee},
  {Owen}, {Buzasi}, {Charbonneau}, {Doyle}, {Fortney}, {Ford}, {Holman},
  {Seager}, {Steffen}, {Welsh}, {Rowe}, {Anderson}, {Buchhave}, {Ciardi},
  {Walkowicz}, {Sherry}, {Horch}, {Isaacson}, {Everett}, {Fischer}, {Torres},
  {Johnson}, {Endl}, {MacQueen}, {Bryson}, {Dotson}, {Haas}, {Kolodziejczak},
  {Van Cleve}, {Chandrasekaran}, {Twicken}, {Quintana}, {Clarke}, {Allen},
  {Li}, {Wu}, {Tenenbaum}, {Verner}, {Bruhweiler}, {Barnes}, \&
  {Prsa}}]{2010Sci...327..977B}
{Borucki}, W.~J., {Koch}, D., {Basri}, G., {et~al.} 2010, Science, 327, 977

\bibitem[{{Borucki} {et~al.}(2011){Borucki}, {Koch}, \&
  {Basri}}]{2011ApJ...736...19B}
{Borucki}, W.~J., {Koch}, D.~G., \& {Basri}, G., e.~a. 2011, \apj, 736, 19

\bibitem[{{Bouchy} {et~al.}(2009){Bouchy}, {H{\'e}brard}, \&
  {Udry}}]{2009A&A...505..853B}
{Bouchy}, F., {H{\'e}brard}, G., \& {Udry}, S., e.~a. 2009, \aap, 505, 853

\bibitem[{{Burrows} {et~al.}(2007){Burrows}, {Hubeny}, \&
  {Budaj}}]{2007ApJ...661..502B}
{Burrows}, A., {Hubeny}, I., \& {Budaj}, J., e.~a. 2007, \apj, 661, 502

\bibitem[{{Claret}(2000)}]{2000A&A...363.1081C}
{Claret}, A. 2000, \aap, 363, 1081

\bibitem[{{Claret}(2004)}]{2004A&A...428.1001C}
{Claret}, A. 2004, \aap, 428, 1001

\bibitem[{{Collier Cameron} {et~al.}(2006){Collier Cameron}, {Pollacco}, \&
  {Street}}]{2006MNRAS.373..799C}
{Collier Cameron}, A., {Pollacco}, D., \& {Street}, R.~A., e.~a. 2006, \mnras,
  373, 799

\bibitem[{{Collier Cameron} {et~al.}(2007){Collier Cameron}, {Wilson}, \&
  {West}}]{2007MNRAS.380.1230C}
{Collier Cameron}, A., {Wilson}, D.~M., \& {West}, R.~G., e.~a. 2007, \mnras,
  380, 1230

\bibitem[{{Doyle} {et~al.}(2014){Doyle}, {Davies}, \&
  {Smalley}}]{2014MNRAS.444.3592D}
{Doyle}, A.~P., {Davies}, G.~R., \& {Smalley}, B., e.~a. 2014, \mnras, 444,
  3592

\bibitem[{{Doyle} {et~al.}(2013){Doyle}, {Smalley}, \&
  {Maxted}}]{2013MNRAS.428.3164D}
{Doyle}, A.~P., {Smalley}, B., \& {Maxted}, P.~F.~L., e.~a. 2013, \mnras, 428,
  3164

\bibitem[{{Dressing} \& {Charbonneau}(2013)}]{2013ApJ...767...95D}
{Dressing}, C.~D. \& {Charbonneau}, D. 2013, \apj, 767, 95

\bibitem[{{Enoch} {et~al.}(2011){Enoch}, {Anderson}, \&
  {Barros}}]{2011AJ....142...86E}
{Enoch}, B., {Anderson}, D.~R., \& {Barros}, S.~C.~C., e.~a. 2011, \aj, 142, 86

\bibitem[{{Fabrycky} \& {Tremaine}(2007)}]{2007ApJ...669.1298F}
{Fabrycky}, D. \& {Tremaine}, S. 2007, \apj, 669, 1298

\bibitem[{{Faedi} {et~al.}(2011){Faedi}, {Barros}, \&
  {Anderson}}]{2011A&A...531A..40F}
{Faedi}, F., {Barros}, S.~C.~C., \& {Anderson}, D.~R., e.~a. 2011, \aap, 531,
  A40

\bibitem[{{Faedi} {et~al.}(2013){Faedi}, {Pollacco}, \&
  {Barros}}]{2013A&A...551A..73F}
{Faedi}, F., {Pollacco}, D., \& {Barros}, S.~C.~C., e.~a. 2013, \aap, 551, A73

\bibitem[{{Ford} \& {Rasio}(2008)}]{2008ApJ...686..621F}
{Ford}, E.~B. \& {Rasio}, F.~A. 2008, \apj, 686, 621

\bibitem[{{Fortney} {et~al.}(2007){Fortney}, {Marley}, \&
  {Barnes}}]{2007ApJ...659.1661F}
{Fortney}, J.~J., {Marley}, M.~S., \& {Barnes}, J.~W. 2007, \apj, 659, 1661

\bibitem[{{Fressin} {et~al.}(2013){Fressin}, {Torres}, {Charbonneau}, {Bryson},
  {Christiansen}, {Dressing}, {Jenkins}, {Walkowicz}, \&
  {Batalha}}]{2013ApJ...766...81F}
{Fressin}, F., {Torres}, G., {Charbonneau}, D., {et~al.} 2013, \apj, 766, 81

\bibitem[{{Gillon} {et~al.}(2013){Gillon}, {Anderson}, \&
  {Collier-Cameron}}]{2013A&A...552A..82G}
{Gillon}, M., {Anderson}, D.~R., \& {Collier-Cameron}, A., e.~a. 2013, \aap,
  552, A82

\bibitem[{{Gillon} {et~al.}(2011){Gillon}, {Jehin}, {Magain}, {Chantry},
  {Hutsem{\'e}kers}, {Manfroid}, {Queloz}, \& {Udry}}]{2011EPJWC..1106002G}
{Gillon}, M., {Jehin}, E., {Magain}, P., {et~al.} 2011, in European Physical
  Journal Web of Conferences, Vol.~11, European Physical Journal Web of
  Conferences, 06002

\bibitem[{{Grunblatt} {et~al.}(2016){Grunblatt}, {Huber}, {Gaidos}, {Lopez},
  {Fulton}, {Fortney}, {Howard}, {Sinukoff}, {Mann}, \&
  {Isaacson}}]{2016arXiv160605818G}
{Grunblatt}, S.~K., {Huber}, D., {Gaidos}, E.~J., {et~al.} 2016, ArXiv e-prints
  [\eprint[arXiv]{1606.05818}]

\bibitem[{{Guillot} {et~al.}(1996){Guillot}, {Burrows}, \&
  {Hubbard}}]{1996ApJ...459L..35G}
{Guillot}, T., {Burrows}, A., \& {Hubbard}, W.~B., e.~a. 1996, \apjl, 459, L35

\bibitem[{{Hartman} {et~al.}(2011){Hartman}, {Bakos}, \&
  {Kipping}}]{2011ApJ...728..138H}
{Hartman}, J.~D., {Bakos}, G.~{\'A}., \& {Kipping}, D.~M., e.~a. 2011, \apj,
  728, 138

\bibitem[{{Hastings}(1970)}]{Hastings70}
{Hastings}, W.~K. 1970, Biometrika, 57, 97

\bibitem[{{H{\'e}brard} {et~al.}(2013){H{\'e}brard}, {Collier Cameron}, \&
  {Brown}}]{2013A&A...549A.134H}
{H{\'e}brard}, G., {Collier Cameron}, A., \& {Brown}, D.~J.~A., e.~a. 2013,
  \aap, 549, A134

\bibitem[{{Hellier} {et~al.}(2016){Hellier}, {Anderson}, \& {Collier
  Cameron}}]{2016arXiv160404195H}
{Hellier}, C., {Anderson}, D.~R., \& {Collier Cameron}, A., e.~a. 2016, ArXiv
  e-prints [\eprint[arXiv]{1604.04195}]

\bibitem[{{Howell} {et~al.}(2014){Howell}, {Sobeck}, \&
  {Haas}}]{2014PASP..126..398H}
{Howell}, S.~B., {Sobeck}, C., \& {Haas}, M., e.~a. 2014, \pasp, 126, 398

\bibitem[{{Huang} \& {Cumming}(2012)}]{2012ApJ...757...47H}
{Huang}, X. \& {Cumming}, A. 2012, \apj, 757, 47

\bibitem[{{Jehin} {et~al.}(2011){Jehin}, {Gillon}, \&
  {Queloz}}]{2011Msngr.145....2J}
{Jehin}, E., {Gillon}, M., \& {Queloz}, D., e.~a. 2011, The Messenger, 145, 2

\bibitem[{{Kov{\'a}cs} {et~al.}(2002){Kov{\'a}cs}, {Zucker}, \&
  {Mazeh}}]{2002A&A...391..369K}
{Kov{\'a}cs}, G., {Zucker}, S., \& {Mazeh}, T. 2002, \aap, 391, 369

\bibitem[{{Kozai}(1962)}]{1962AJ.....67..591K}
{Kozai}, Y. 1962, \aj, 67, 591

\bibitem[{{Law} {et~al.}(2015){Law}, {Fors}, {Ratzloff}, {Wulfken},
  {Kavanaugh}, {Sitar}, {Pruett}, {Birchard}, {Barlow}, {Cannon}, {Cenko},
  {Dunlap}, {Kraus}, \& {Maccarone}}]{2015PASP..127..234L}
{Law}, N.~M., {Fors}, O., {Ratzloff}, J., {et~al.} 2015, \pasp, 127, 234

\bibitem[{{Lendl} {et~al.}(2012){Lendl}, {Anderson}, \&
  {Collier-Cameron}}]{2012A&A...544A..72L}
{Lendl}, M., {Anderson}, D.~R., \& {Collier-Cameron}, A., e.~a. 2012, \aap,
  544, A72

\bibitem[{{Lidov}(1962)}]{1962P&SS....9..719L}
{Lidov}, M.~L. 1962, \planss, 9, 719

\bibitem[{{Lopez} \& {Fortney}(2016)}]{2016ApJ...818....4L}
{Lopez}, E.~D. \& {Fortney}, J.~J. 2016, \apj, 818, 4

\bibitem[{{Mancini} {et~al.}(2015){Mancini}, {Esposito}, \&
  {Covino}}]{2015A&A...579A.136M}
{Mancini}, L., {Esposito}, M., \& {Covino}, E., e.~a. 2015, \aap, 579, A136

\bibitem[{{Mandel} \& {Agol}(2002)}]{2002ApJ...580L.171M}
{Mandel}, K. \& {Agol}, E. 2002, \apjl, 580, L171

\bibitem[{{Matsumura} {et~al.}(2010){Matsumura}, {Peale}, \&
  {Rasio}}]{2010ApJ...725.1995M}
{Matsumura}, S., {Peale}, S.~J., \& {Rasio}, F.~A. 2010, \apj, 725, 1995

\bibitem[{{Maxted} {et~al.}(2011){Maxted}, {Anderson}, {Collier Cameron},
  {Hellier}, {Queloz}, {Smalley}, {Street}, {Triaud}, {West}, {Gillon},
  {Lister}, {Pepe}, {Pollacco}, {S{\'e}gransan}, {Smith}, \&
  {Udry}}]{2011PASP..123..547M}
{Maxted}, P.~F.~L., {Anderson}, D.~R., {Collier Cameron}, A., {et~al.} 2011,
  \pasp, 123, 547

\bibitem[{{Maxted} {et~al.}(2015{\natexlab{a}}){Maxted}, {Serenelli}, \&
  {Southworth}}]{2015A&A...575A..36M}
{Maxted}, P.~F.~L., {Serenelli}, A.~M., \& {Southworth}, J. 2015{\natexlab{a}},
  \aap, 575, A36

\bibitem[{{Maxted} {et~al.}(2015{\natexlab{b}}){Maxted}, {Serenelli}, \&
  {Southworth}}]{2015A&A...577A..90M}
{Maxted}, P.~F.~L., {Serenelli}, A.~M., \& {Southworth}, J. 2015{\natexlab{b}},
  \aap, 577, A90

\bibitem[{{Mazeh} {et~al.}(2016){Mazeh}, {Holczer}, \&
  {Faigler}}]{2016A&A...589A..75M}
{Mazeh}, T., {Holczer}, T., \& {Faigler}, S. 2016, \aap, 589, A75

\bibitem[{{Metropolis} {et~al.}(1953){Metropolis}, {Rosenbluth}, {Rosenbluth},
  {Teller}, \& {Teller}}]{1953JChPh..21.1087M}
{Metropolis}, N., {Rosenbluth}, A.~W., {Rosenbluth}, M.~N., {Teller}, A.~H., \&
  {Teller}, E. 1953, \jcp, 21, 1087

\bibitem[{{P{\'a}l} {et~al.}(2016){P{\'a}l}, {M{\'e}sz{\'a}ros}, {Jask{\'o}},
  {Mez{\H o}}, {Cs{\'e}p{\'a}ny}, {Vida}, \& {Ol{\'a}h}}]{2016PASP..128d5002P}
{P{\'a}l}, A., {M{\'e}sz{\'a}ros}, L., {Jask{\'o}}, A., {et~al.} 2016, \pasp,
  128, 045002

\bibitem[{{Pepe} {et~al.}(2002){Pepe}, {Mayor}, \&
  {Galland}}]{2002A&A...388..632P}
{Pepe}, F., {Mayor}, M., \& {Galland}, F., e.~a. 2002, \aap, 388, 632

\bibitem[{{Pepper} {et~al.}(2007){Pepper}, {Pogge}, {DePoy}, {Marshall},
  {Stanek}, {Stutz}, {Poindexter}, {Siverd}, {O'Brien}, {Trueblood}, \&
  {Trueblood}}]{2007PASP..119..923P}
{Pepper}, J., {Pogge}, R.~W., {DePoy}, D.~L., {et~al.} 2007, \pasp, 119, 923

\bibitem[{{Petigura} {et~al.}(2013){Petigura}, {Howard}, \&
  {Marcy}}]{2013PNAS..11019273P}
{Petigura}, E.~A., {Howard}, A.~W., \& {Marcy}, G.~W. 2013, Proceedings of the
  National Academy of Science, 110, 19273

\bibitem[{{Pollacco} {et~al.}(2008){Pollacco}, {Skillen}, \& {Collier
  Cameron}}]{2008MNRAS.385.1576P}
{Pollacco}, D., {Skillen}, I., \& {Collier Cameron}, A., e.~a. 2008, \mnras,
  385, 1576

\bibitem[{{Pollacco} {et~al.}(2006){Pollacco}, {Skillen}, \& {Collier
  Cameron}}]{2006PASP..118.1407P}
{Pollacco}, D.~L., {Skillen}, I., \& {Collier Cameron}, A., e.~a. 2006, \pasp,
  118, 1407

\bibitem[{{Queloz} {et~al.}(2000){Queloz}, {Eggenberger}, \&
  {Mayor}}]{2000A&A...359L..13Q}
{Queloz}, D., {Eggenberger}, A., \& {Mayor}, M., e.~a. 2000, \aap, 359, L13

\bibitem[{{Queloz} {et~al.}(2001){Queloz}, {Henry}, \&
  {Sivan}}]{2001A&A...379..279Q}
{Queloz}, D., {Henry}, G.~W., \& {Sivan}, J.~P., e.~a. 2001, \aap, 379, 279

\bibitem[{{Ricker} {et~al.}(2015){Ricker}, {Winn}, {Vanderspek}, {Latham},
  {Bakos}, {Bean}, {Berta-Thompson}, {Brown}, {Buchhave}, {Butler}, {Butler},
  {Chaplin}, {Charbonneau}, {Christensen-Dalsgaard}, {Clampin}, {Deming},
  {Doty}, {De Lee}, {Dressing}, {Dunham}, {Endl}, {Fressin}, {Ge}, {Henning},
  {Holman}, {Howard}, {Ida}, {Jenkins}, {Jernigan}, {Johnson}, {Kaltenegger},
  {Kawai}, {Kjeldsen}, {Laughlin}, {Levine}, {Lin}, {Lissauer}, {MacQueen},
  {Marcy}, {McCullough}, {Morton}, {Narita}, {Paegert}, {Palle}, {Pepe},
  {Pepper}, {Quirrenbach}, {Rinehart}, {Sasselov}, {Sato}, {Seager},
  {Sozzetti}, {Stassun}, {Sullivan}, {Szentgyorgyi}, {Torres}, {Udry}, \&
  {Villasenor}}]{2015JATIS...1a4003R}
{Ricker}, G.~R., {Winn}, J.~N., {Vanderspek}, R., {et~al.} 2015, Journal of
  Astronomical Telescopes, Instruments, and Systems, 1, 014003

\bibitem[{{Schlaufman} \& {Winn}(2013)}]{2013ApJ...772..143S}
{Schlaufman}, K.~C. \& {Winn}, J.~N. 2013, \apj, 772, 143

\bibitem[{{Smalley} {et~al.}(2012){Smalley}, {Anderson}, \&
  {Collier-Cameron}}]{2012A&A...547A..61S}
{Smalley}, B., {Anderson}, D.~R., \& {Collier-Cameron}, A., e.~a. 2012, \aap,
  547, A61

\bibitem[{{Snellen} {et~al.}(2012){Snellen}, {Stuik}, {Navarro}, {Bettonvil},
  {Kenworthy}, {de Mooij}, {Otten}, {ter Horst}, \& {le
  Poole}}]{2012SPIE.8444E..0IS}
{Snellen}, I.~A.~G., {Stuik}, R., {Navarro}, R., {et~al.} 2012, in \procspie,
  Vol. 8444, Ground-based and Airborne Telescopes IV, 84440I

\bibitem[{{Southworth} {et~al.}(2014){Southworth}, {Hinse}, \&
  {Burgdorf}}]{2014MNRAS.444..776S}
{Southworth}, J., {Hinse}, T.~C., \& {Burgdorf}, M., e.~a. 2014, \mnras, 444,
  776

\bibitem[{{Southworth} {et~al.}(2012){Southworth}, {Hinse}, {Dominik}, {Fang},
  {Harps{\o}e}, {J{\o}rgensen}, {Kerins}, {Liebig}, {Mancini}, {Skottfelt},
  {Anderson}, {Smalley}, {Tregloan-Reed}, {Wertz}, {Alsubai}, {Bozza}, {Calchi
  Novati}, {Dreizler}, {Gu}, {Hundertmark}, {Jessen-Hansen}, {Kains},
  {Kjeldsen}, {Lund}, {Lundkvist}, {Mathiasen}, {Penny}, {Rahvar}, {Ricci},
  {Scarpetta}, {Snodgrass}, \& {Surdej}}]{2012MNRAS.426.1338S}
{Southworth}, J., {Hinse}, T.~C., {Dominik}, M., {et~al.} 2012, \mnras, 426,
  1338

\bibitem[{{Steele} {et~al.}(2008){Steele}, {Bates}, \&
  {Gibson}}]{2008SPIE.7014E..6JS}
{Steele}, I.~A., {Bates}, S.~D., \& {Gibson}, N., e.~a. 2008, in \procspie,
  Vol. 7014, Ground-based and Airborne Instrumentation for Astronomy II, 70146J

\bibitem[{{Stetson}(1987)}]{1987PASP...99..191S}
{Stetson}, P.~B. 1987, \pasp, 99, 191

\bibitem[{{Tamuz} {et~al.}(2005){Tamuz}, {Mazeh}, \&
  {Zucker}}]{2005MNRAS.356.1466T}
{Tamuz}, O., {Mazeh}, T., \& {Zucker}, S. 2005, \mnras, 356, 1466

\bibitem[{{van Saders} {et~al.}(2016){van Saders}, {Ceillier}, {Metcalfe},
  {Silva Aguirre}, {Pinsonneault}, {Garc{\'{\i}}a}, {Mathur}, \&
  {Davies}}]{2016Natur.529..181V}
{van Saders}, J.~L., {Ceillier}, T., {Metcalfe}, T.~S., {et~al.} 2016, \nat,
  529, 181

\bibitem[{{Villaver} \& {Livio}(2009)}]{2009ApJ...705L..81V}
{Villaver}, E. \& {Livio}, M. 2009, \apjl, 705, L81

\bibitem[{{Weiss} \& {Schlattl}(2008)}]{2008Ap&SS.316...99W}
{Weiss}, A. \& {Schlattl}, H. 2008, \apss, 316, 99

\bibitem[{{Wheatley}(2013)}]{2013EPSC....8..234W}
{Wheatley}, P. 2013, European Planetary Science Congress 2013, held 8-13
  September in London, UK.~Online at: <A
  href=''http://meetings.copernicus.org/epsc2013''>
  http://meetings.copernicus.org/epsc2013</A>, id.EPSC2013-234, 8, EPSC2013

\end{thebibliography}
%\bibliography{}

\end{document}